\documentclass[AMA,STIX1COL]{WileyNJD-v2}

\usepackage{bigstrut}

\usepackage{tikz}

\newcommand\copyrighttext{%
  \footnotesize This work has been submitted to the International Journal of Network Management (IJNM) and accepted for publication. Copyright may be transferred without notice, after which this version may no longer be accessible.}
\newcommand\copyrightnotice{%
\begin{tikzpicture}[remember picture,overlay]
\node[anchor=south,yshift=10pt] at (current page.south) {\fbox{\parbox{\dimexpr\textwidth-\fboxsep-\fboxrule\relax}{\copyrighttext}}};
\end{tikzpicture}%
}

\articletype{Article Type}

\received{26 April 2016}
\revised{6 June 2016}
\accepted{6 June 2016}

\begin{document}

\title{An Access Control for IoT Based on Network Community Perception and Social Trust Against Sybil Attacks}

\author[1]{Gustavo Oliveira*}

\author[1]{Agnaldo de Souza Batista}

\author[1,2]{Michele Nogueira}

\author[1,2]{Aldri Santos}

\authormark{OLIVEIRA \textsc{et al}}

\address[1]{\orgdiv{Department of Informatics}, \orgname{Federal University of Paraná}, \orgaddress{\state{Paraná}, \country{Brazil}}}

\address[2]{\orgdiv{Department of Computer Science}, \orgname{Federal University of Minas Gerais}, \orgaddress{\state{Minas Gerais}, \country{Brazil}}}

\corres{Aldri Santos, Department of Computer Science, Federal University of Minas Gerais,  Brazil. Email: aldri@dcc.ufmg.br}

\abstract[Summary]{The evolution of the Internet of Things (IoT) has increased the connection of personal devices, 
mainly taking into account the habits and behavior of their owners. These environments demand access control mechanisms to protect them against intruders, like Sybil attacks. that can compromise data privacy or disrupt the network operation. The Social IoT paradigm enables access control systems to aggregate community context and sociability information from devices to enhance robustness and security. This work introduces the \mbox{ELECTRON} mechanism to control access in IoT networks based on social trust between devices to protect the network from Sybil attackers. \mbox{ELECTRON} groups IoT devices into communities by their social similarity and evaluates their social trust, strengthening the reliability between legitimate devices and their resilience against the interaction of Sybil attackers. NS-3 Simulations show the \mbox{ELECTRON} performance under Sybil attacks on several IoT communities so that it has gotten to detect more than 90\% of attackers in a scenario with 150 nodes into offices, schools, gyms, and~parks communities, and in other scenarios for same communities it achieved around of 90\% of detection. Furthermore, it provided high accuracy, over 90-95\%, and false positive rates closer to zero.}


\jnlcitation{\cname{
\author{A. Santos}}, 
\author{G. Oliveira}, 
\author{A. Batista}, and
\author{M. Nogueira}, (\cyear{2021}), 
\ctitle{An Access Control for IoT Based on Network Community Perception and Social Trust Against Sybil Attacks}}

\maketitle
\copyrightnotice

\section{Introduction}\label{sec:intro}

The Internet of Things (IoT) has improved communication between computational devices and people objects~\cite{zeadally2019cryptographic} and allowed several domains and applications to promote the existence of services that contribute to citizens' quality of life. The hyper-connectivity of the IoT enables electronic sensors and sensor systems to become huge generators of data, currently reaching yearly rates on the zettabyte scale.~\cite{Reinsel2018,erhan2020smart} Hence, data dissemination is essential to ensure the availability of services, since it allows the operation of many applications in the IoT ecosystem, such as environmental sensing, monitoring of vital signs, among others.~\cite{Evangelista2016IEEE} Moreover, data dissemination is one of the most affected services in IoT, which points out the relevance of data security and privacy to the future of this network.~\cite{Sicari2015} Among other factors, security is one of the most critical factors for obtaining a secure system and environment. Additionally, ensuring security can avoid serious risks.~\cite{lee2021survey}

The data dissemination and sharing in IoT naturally demand an effective service of access control (AC), adaptive to changes in the context, and resilient to non-authorized network access.~\cite{castro2019casa, antunes2018maniot} However, a couple of issues inherent to IoT (e.g., heterogeneity, mobility, computing resource, and power) often inhibit the use of such services.~\cite{santos2019clustering} At the system-level, the maintenance of system operation availability relies on security methods~\cite{cervantes2015detection} to analyze physical behavior and support the decision-making,~\cite{clemente2018cyber} as the threat of attacks is always a possibility, demanding measures to detect them and mitigate their effects.~\cite{jahromi2021toward, pelloso2018self}. In this way, attacks like Sybil, for instance, exploit these issues to compromise the privacy and reliability of the disseminated data.~\cite{Pongle2015} A Sybil attacker aims to imitate a legitimate user, forging or stealing identities to a computational device. Once it gains access to the network, the users' data become available to it, as well as increases the number of control messages due to this misbehavior.~ \cite{Medjek2017} Although classic access control techniques such as Role-Based Access Control (RBAC)~\cite{Ferraiolo1995} and Attribute-Based Access Control (ABAC)~\cite{Yuan2005, Kuhn2010} bring up network scalability, they add extra overhead in the network,~\cite{Gusmeroli2012} which inhibits their use in IoT networks with a large number of connected devices. Meanwhile, techniques like Capacity-Based Access Control (CapBAC), Risk-Based Access Control (RiskBAC), and Trust-Based Access Control (TBAC) are more resilient and adaptive, which makes them suitable to IoT.~\cite{Ouaddah2017} On the CapBAC model, all devices access the context and services through keys distributed by a certificate authority,~\cite{Hernandez2016,Hussein2017} whereas the RBAC model focuses on estimating vulnerabilities and threats to each requested access by evidences and input parameters.~\cite{Alenezi2017} Lastly, the TBAC model concentrates on computing the device reliability based on its experiences (i.e., direct trust) and neighborhood recommendations techniques.~\cite{Yan2014} Recently, some dynamic access control techniques aim to face issues related to the dynamic nature of IoT, such as the Dynamic Access Control Framework for the IoT (DACIoT). This model provide automatic policies generation, continuous policy enforcement and adaptive policies adjustments.~\cite{alkhresheh2020daciot}

Despite the advances in security solutions, they still face scalability, centralization, and unclear design issues. Furthermore, they rely on hardware features and disregard the social intelligence that arises from the relationships among the devices and owners. Hence, such solutions do not enhance real interactions between individuals, where trust comes from human relationships in communities like family, workplaces, and friendships, for instance. A suitable approach to deal with those challenges comes from social relations, as they make it possible to build higher trust levels to control access to places and private personal and device data.~\cite{Abderrahim2017} The use of subjective aspects from social relationship can ensure reliable data analysis, qualified services and enhanced security in the Social Internet~of Things (SIoT) paradigm.~\cite{abdelghani2016trust} Thus, the association of the devices relies on human relationships in order to propagate context on~the network.

This article introduces the \mbox{ELECTRON} ({\it Acc{\bf{\it E}}ss Contro{\bf {\it L}} Driv{\bf {\it E}}n on {\bf {\it C}}o\-mmu\-nity and Social {\bf {\it TR}}ust {\bf {\it O}}f Thi{\bf {\it N}}gs}) mechanism in order to support the access control on IoT networks against Sybil attacks. It relies on devices social trust established by their sociability perceptions, so that communities formed by devices social similarity make it possible to detect Sybil attackers.  \mbox{ELECTRON} plays in a distributed way, being hence scalable, context-sensitive, and energetically feasible. NS-3 Simulation results demonstrate the  \mbox{ELECTRON} effectiveness against Sybil attacks on IoT. In general, \mbox{ELECTRON} detected more than 90\% of attackers made with stolen or fabricated identities for communities established into offices, schools, gyms and parks, in a scenario with 150 nodes. For scenarios with 100 and 200 nodes, it detected about of 90\% of attacks on the same communities. It also achieved high accuracy, over 90-95\% in specific scenarios, and false positive rates closer to zero due to social relations provided by device interactions and their similarity into within available communities. Therefore, \mbox{ELECTRON} has proven to be adaptable and flexible since it has been evaluated in distinct environments and different communities.  

The remainder of the article is structured as follows: Section~\ref{sec:related} presents a brief background and the related work, Section~\ref{sec:proposal} describes the \mbox{ELECTRON} mechanism, detailing how it controls the access to the network and detects Sybil attackers. Section~\ref{sec:eval} details an comparative analysis of the \mbox{ELECTRON} against a target tracking approach. Lastly, Section~\ref{sec:concl} concludes the work.

\section{Background and Related Work}\label{sec:related}

This section presents an overview of the Social Internet of Things (SIoT) paradigm, highlighting the type of social relations that support it. Next, we take place a discussion about the related work bringing some access control solutions to IoT environments. Lastly, we explain the behaviors of Sybil attacks and how they compromise data privacy or disrupt the network operation.

The possibility for objects to create and manage social relations contributed to the development of several services inside the Internet of Things (IoT), enabling data dissemination, service discovery, selection, and composition.~\cite{Atzori2011,Atzori2012} Integrating social networking concepts into the IoT has led to the SIoT paradigm, which enables people and connected devices to interact, facilitating information sharing and enabling a variety of attractive applications,~\cite{abdelghani2016trust} as long as to establish trust indicators based on these interactions in the IoT environment.~\cite{Truong2017} SIoT aims to bring human social relations to objects so that they can mimic them. The creation and management of these relationships happen without human intervention, generating an artificial social intelligence between the objects. Further, as trust is a fundamental aspect of human social relations, its application in the SIoT environment relies on a trustee's perspective over a set of individuals.

\begin{figure}[!ht]
    \centering
    \includegraphics[width=.5\textwidth]{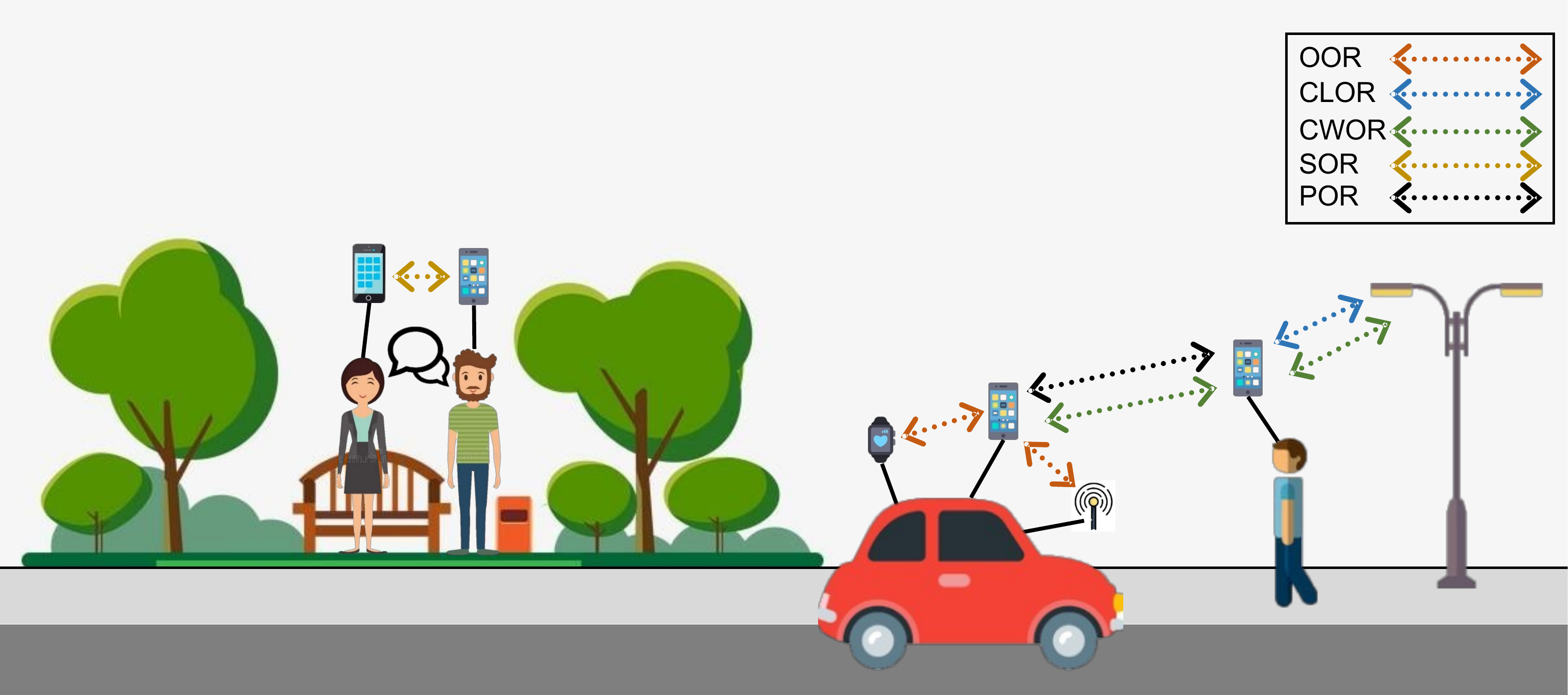}
    \caption{The Type of SIoT Relationships}
    \label{fig:SIoTRelations}
\end{figure}

The SIoT model envisions distinct object relations. According to Atzori et al.,~\cite{Atzori2012} objects can establish five types of relationships, as depicts Figure~\ref{fig:SIoTRelations}. The first one is the {\it parental object relationship} (POR), which is created between objects fabricated at the same time and by the same manufacturer, and is commonly established by homogeneous objects. Next, there is the {\it ownership object relationship} (OOR), which occurs between heterogeneous objects belonging to the same owner, such as smartphones, tablets, and laptops. Another one is the {\it co-location object relationship} (CLOR), which happens between homogeneous and heterogeneous objects always located in the same place, like sensors, actuators, and others objects. When objects collaborate to prove a common application, they establish a {\it co-work object relationship} (CWOR). Finally, there is the {\it social object relationship} (SOR), which comes from object owners' relations and arises when their objects interact with each other.

The main idea behind these relations is that objects with similar characteristics can share their experiences for solving common problems, just as they were friends. Therefore, it is expected that these relationships are responsible for generating a set of rules, social interactions, and services for the objects. By following approaches inspired by human social relations and relationship models, the objects aggregate human behavior to obtain better effectiveness in the services provided, extending to the object community the models and principles of the studies in human social networks. All those relationships play a key role in creating more scalable and secure models for IoT. Several works employ social relations between objects to trust evaluation,~\cite{Bernabe2016,Jayasinghe2016} but in this work, we rely on social relations as support to mechanism operation by extracting the previously described benefits from SIoT to integrate a more efficient access control service.

The ecosystem of IoT devices fosters the existence of several services, and most of them demand suitable protection to avoid the disruption of data privacy and integrity.~\cite{Greengard2019} Recently, the IoT security problems are increasing, requiring recognizing the anomalies attempting to identify data patterns that deviate remarkably from the expected behavior.~\cite{erhan2020smart} Notably, the access control service enhances the security in the network by creating the first barrier through authentication mechanisms, since authentication identifies users and devices in a network and grants access to authorized persons and non manipulated devices.~\cite{hassan2019current} However, devices with constrained resources in IoT inhibit the use of traditional techniques. Although there are several solutions to provide access control, like ones based on capability (CapBAC),  risk (RiskBAC) and  trust (TBAC), among others, they are commonly suitable to environments with heterogeneous devices with constrained resources, naturally found in IoT, mainly due to their mobility and lower resource demand. Moreover, technologies such as blockchain also have been applied to support control access mechanisms based on smart contracts.

In the CapBAC model, the access privileges of a device rely on a key under control of a certificate authority. Anggorojati et al.~\cite{Anggorojati2012} presented a context-aware access control based on capability that combines context information (i.e., local and time) to the traditional CapBAC model to achieve scalability and flexibility. Thereafter, they optimized the model adding cryptography by elliptic curves to operate in a distributed fashion.~\cite{Hernandez2016} On the other hand, the Distributed Capability-based Access Control (DCapBAC) approach meets issues like scalability, interoperability, and end-to-end security, as it works in a distributed way and employs cryptography techniques like Public Key Cryptography (PKC) and Elliptic Curve Cryptography (ECC). But, DCapBAC disregards issues related to granularity and context awareness. The absence of details about the access control initialization procedure and the generation of the access keys difficult the approach verification on the network. Further, there is the need of trust between domains, which demands authentication mechanisms and signature entities.

A Community Capability-Based Access Control (COCapBAC) framework, presented by Hussein et al.,~\cite{Hussein2017} incorporates a community-based structure to support the access control on a distributed environment. Only smart objects that~share the same roles establish communities, what do not follow the SIoT paradigm. Hence, devices with more resources make decisions about the access to devices with constrained resources. However, due to high mobility and intense social interactions of some devices, the community concept lifts up challenges about the absence of autonomy of objects to apply its features like similarity to support a secure and individual access evaluation. Further, it is not analyzed the impact of building new access keys.

We can classify risk-based approaches into adaptive and non-adaptive. While the adaptive ones update the information in real-time to evaluate the risk, non-adaptive ones compute the risk only in their initialization phase. Atlam et al.~\cite{Alenezi2017} presented an Adaptive Risk-Based Access Control (AdRBAC) for IoT based on the risk quantification. Recently, the authors extended their work to adopt the extensible AC markup language (XACML) as a policy language and an architecture for access control in IoT environments.~\cite{atlam2018xacml} The AdRBAC employs factors like context (i.e., place and time), resources sensibility, actions severity, and history to estimate the risk of each access request. Further, it allows for environment heterogeneity and dynamicity. Furthermore, since AdRBAC disregards social aspects of devices, it becomes unsuited to environments where the devices relate to each other.

Recently, the use of blockchain to support AC mechanisms in IoT has been discussed, mainly to leverage some relevant characteristics of this technology to this environment, especially to face security issues. In this context, Zhang et al.~\cite{zhang2018smart} proposed an AC framework for IoT systems based on the Ethereum smart contract.~\cite{ethereum2016} This solution employs blockchain to provide a distributed and trustworthy AC scheme consisting of multiple access control contracts, one judge contract, and one register contract. It offers static access right validation based on predefined access control policies and dynamic access right validation by behavior checking. In this AC model, as each peer in the network may own some resources (e.g., services, data, and storage space) needed by the other peers, resource owners must implement access control to prevent unauthorized use of their resources. The use of blockchain improves security in controlling access to resources. Still, issues related to the time for deploying and running smart contracts in the face of the constrained resources of some IoT devices may limit the framework usage.

The trust concept has been widely discussed in IoT, mainly due to its relevance to enhance network security.~\cite{Gu2014,Chen2014,Nguyen2016,Sato2016,Khan2017} Mahale et al.~\cite{Mahalle2013} presented a Fuzzy approach to the Trust Based Access Control (FTBAC) framework that measures the values of experience, knowledge and recommendations to assess trust. A decision framework relies on fuzzy logic to control the access to the network. Bernabe et al.~\cite{Bernabe2016} added sociability to TBAC model and developed a Trust-aware Access Control System for IoT (TACIoT) to achieve a reliable multidimensional IoT access control aiming to be lightweight, flexible and adaptive. This work turns to an environment with grouped devices by ongoing relationships based on dimensions of assessment of Quality of Service (QoS), security, reputation, and social relationships. Each of these dimensions computes a trust value, and feeds a fuzzy controller that supports the decision making. The trust evaluation relies on social aspects like common interests and friendship. Though, there is a lack of details about the communities construction, and the relation between common interests and friendship throughout the evaluation process. Recently, Sakthivel et al.~\cite{sakthivel2021trust} presented a Trust-Based Access Control Mechanism (Trust ACM) to support users, applications, and devices in various physical locations to easy and seamless communication with each other. Trust ACM computes trust based on direct trust, indirect trust, and reputation. The direct trust relies on packet delivery ratio, availability, and authentication mechanism in the node, ensuring access to a node only by a security mechanism. Indirect trust comes from the recommendation of the neighbor based on its direct interaction with the node. Finally, the reputation is computed using duration through which the node has had high trust and is given as a ratio of time. Trust ACM manages the AC based on trust degrees generated automatically from the trust model, mapping users with their trust degree to reduce the need for manual administration. However, the details about building the presented solution.

The highly dynamic nature of IoT is also an issue acknowledged by works available in the literature. Alkhresheh et al.~\cite{alkhresheh2020daciot} presented a Dynamic Access Control Framework for the Internet of Things (DACIoT), extending the standard reference model of the XACML. Duet to limitations in XACML related to manual access policy management, discontinuous access decision enforcement; and overprovisioning of access permissions, the authors replaced the components Policy Decision Point (PDP), Policy Enforcement Point (PEP), and Policy Administration Point (PAP) with automatic policy specification (APS), continuous policy enforcement (CPE), and adaptive policy adjustment (APA), respectively. DACIoT relies on the association between elements (i.e., objects and persons) and the association of these elements with their context. Although the dynamic of IoT is taken into account, this solution relies on centralized entities to perform the decision-making about the access control.

The literature also presents several techniques to deal with Sybil attacks like the mechanism called \mbox{SA$^{2}$CI} (Sybil Attack Association Control for IoT).~\cite{Evangelista2016IEEE} In this work the authors proposed an association control to prevent Sybil attacks by Elliptic Curve Cryptography (ECC) to safely distribute symmetric keys among network nodes. Further, this mechanism employs Physical Unclonable Functions (PUF) to extract information from devices hardware and associates them with nodes identities to ensure their proof of identity and their non-retractability. Although \mbox{SA$^{2}$CI} identifies Sybil attackers and be energy efficient, it disregards social information, becoming unsuited to environments with high mobility devices.

\subsection{Sybil attacks}

\begin{figure}[!ht]
    \centering
    \includegraphics[width=.35\textwidth]{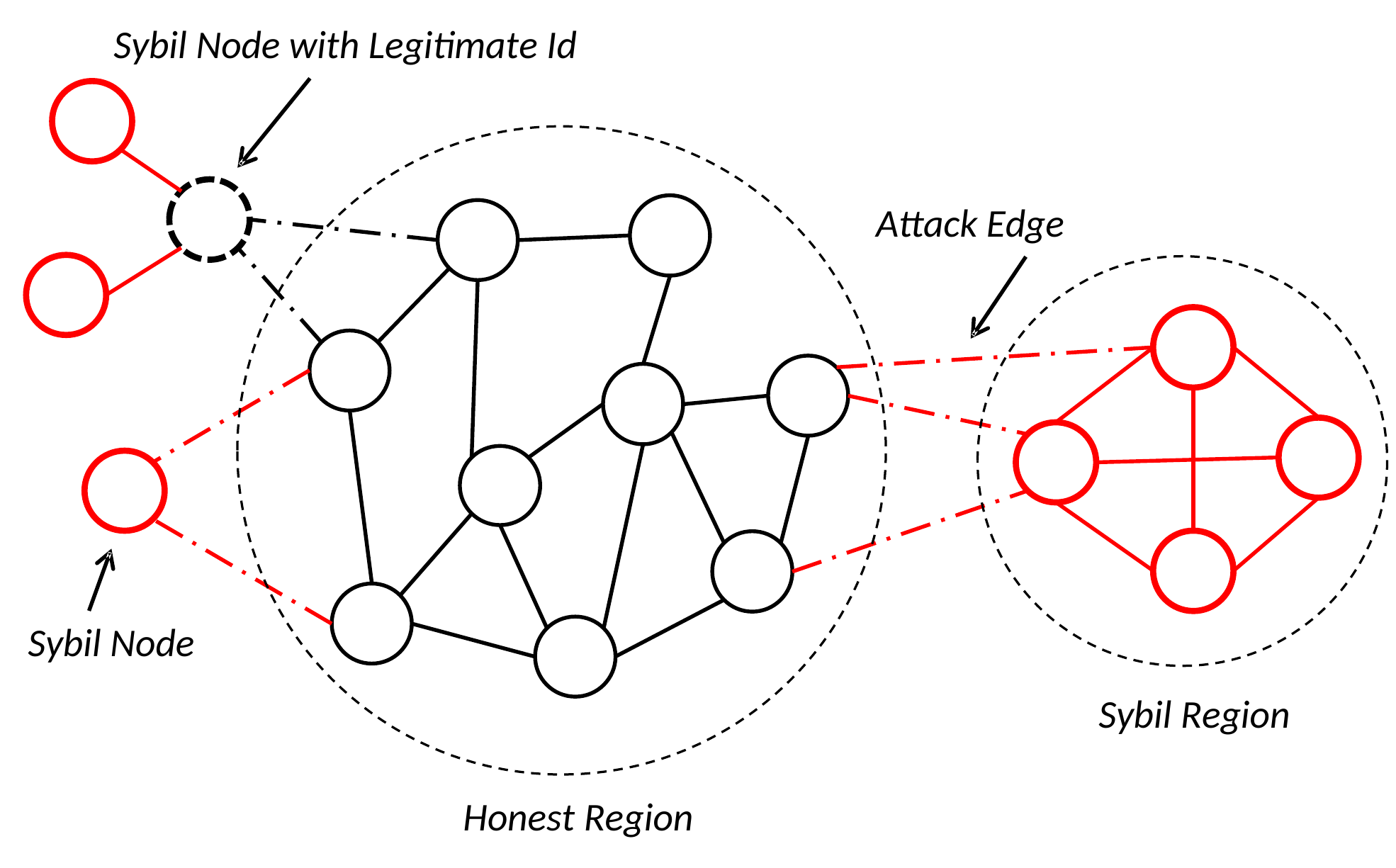}
    \caption{Sybil distinct behaviors}
    \label{fig:TypesSybil}
\end{figure}

A Sybil attack means manipulating false or stolen identities by an attacker eavesdropping on the network and cloning or fabricating identities based on the identities provided to the authentication service. A node acting as a Sybil attacker illegitimately employs multiple identities, thus creating an illusion that many nodes are operating in the network, eventually devastating the topology of the network.~\cite{jan2018sybil} The Sybil attacker focuses on manipulating the network devices, upsetting the communication process,~\cite{Evangelista2016IEEE,bang2021novel} by leaning on communication between legitimate and malicious nodes to launch the attack on a distributed network. Depending on the way the attacker communicates with the network nodes, this attack can happen directly, when the attacker communicates directly with the legitimate nodes, or indirectly, when he/she relies on a legitimate identity to communicate with the legitimate nodes and deliver data to them.~\cite{Valarmathi2016} In a distributed system, a Sybil attacker can employ all Sybil identities simultaneously or in an isolated way, while the majority of the identities remain in an idle state. Essentially, selecting these two schemes depends on how costly it can be for the attacker to obtain identity in the system. When an attacker easily gets many identities, it keeps some of them in an idle state to make the attack seem more real, as the normal behavior of legitimate nodes is to go out and come back into the system many times. Figure~\ref{fig:TypesSybil} depicts some of the Sybil behaviors. In a Sybil region, Sybil nodes join for the attack, while a Sybil node with a legitimate identity injects data into a honest region.

\begin{figure}[!ht]
    \centering
    \includegraphics[width=50mm]{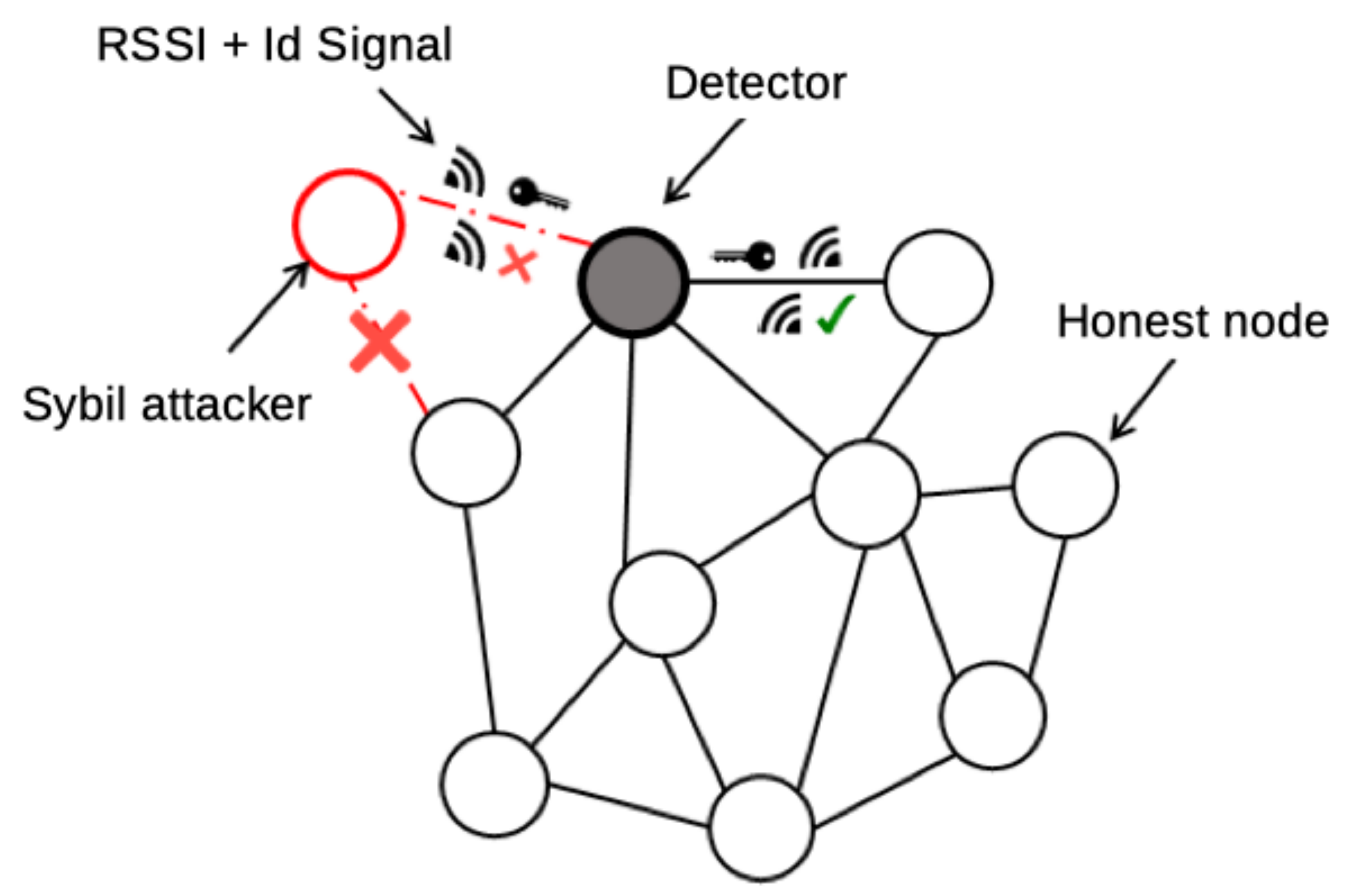}
    ~
    \includegraphics[width=60mm]{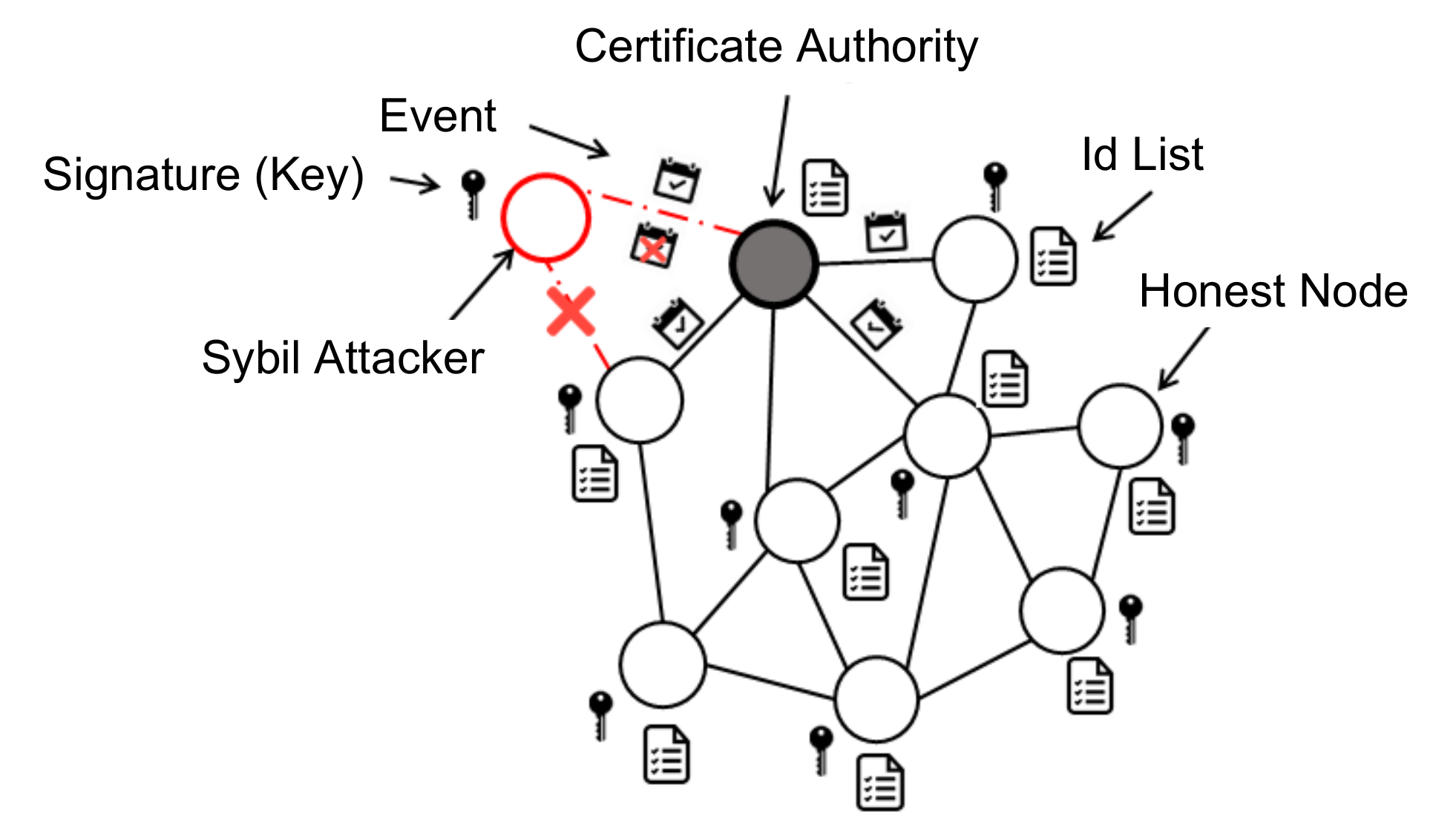}
    ~
    \includegraphics[width=60mm]{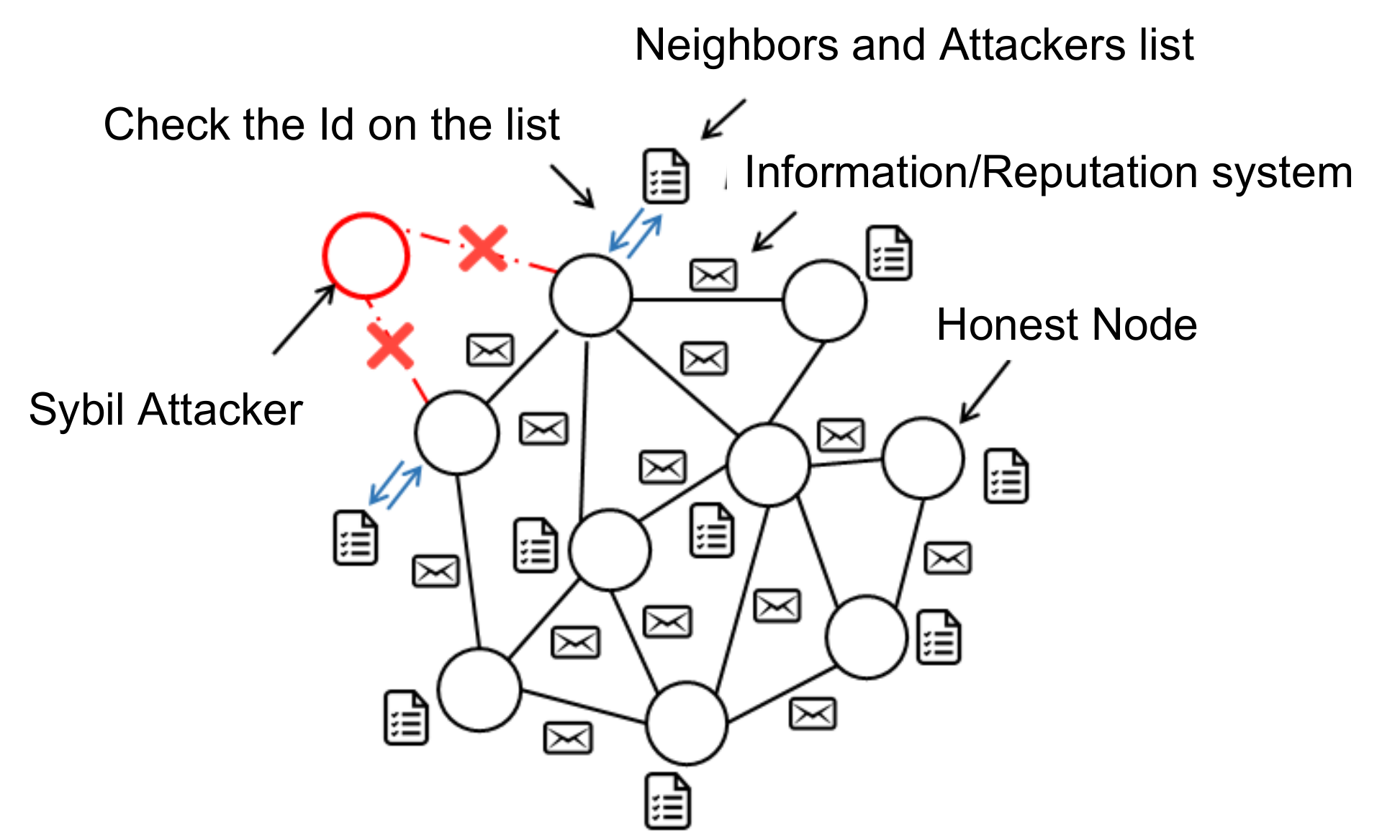}
    \caption{Sybil attacks detection techniques}
    \label{fig:SybilDetection}
\end{figure}

Another possible strategy to a Sybil attack is to simultaneously activate all identities to access the network or activate one by one in a longer attack time. However, as new identities are activated and the attack duration increases, the complexity for the attacker to deal with security mechanisms also increases. Also, the position of the Sybil nodes concerning the network access control collaborates to the success of the attack, as it can make it easy for Sybil nodes to authenticate on the network. When the attacker has at least one real identity acquired from the network, it is inside the network. Otherwise, it is assumed the attacker is out of the network. Currently, techniques for detecting Sybil attacks rely on Received Signal Strength Indication (RSSI), encryption, and reputation systems,~\cite{Evangelista2016IEEE} as shown in Figure.~\ref{fig:SybilDetection} All these techniques offer advantages and disadvantages when applied in IoT.~\cite{Zhang2014} In general, as shows Figure~\ref{fig:SybilDetection} left, RSSI-based approaches consider mobility and coverage area to identify a Sybil behavior on the network.~\cite{Abbas2013} As the mechanisms based on this strategy demand fewer resources, they are suitable to IoT. However, they present low efficiency in Sybil attack detection, generating many false positives.

The approaches based on cryptographic techniques for Sybil attack detection require additional communication and processing, as shows Figure~\ref{fig:SybilDetection} center. These techniques can be asymmetric and symmetric. The generation of a secure asymmetric key demands high processing costs, making it suitable for resourced devices. On the other hand, symmetric keys rely on the key exchange between two nodes to ensure non-repudiation, becoming a bottleneck. Therefore, these approaches imply a trade-off between security and performance, considering resource constraints, scalability, and network overhead.~\cite{Evangelista2016IEEE}  Lastly, as shows Figure~\ref{fig:SybilDetection} right, the reputation-based strategy keeps a list of identities about legitimate and attackers nodes. Nodes build this list by exchanging information about their neighbors' behavior to create a reputation system inside the network. Hence, nodes that do not cooperate have their reputations diminished.~\cite{Quercia2010} Approaches based on this strategy present a high precision in attack detection, but the overload of control messages causes an overhead on the network. Moreover, there is the possibility of the attacker cheating the reputation system with good behavior and the need for offline training of the mechanism to identify malicious behavior. There is always a trade-off between security and performance in all the presented techniques when we select one of the available solutions.

\section{\mbox{ELECTRON}}\label{sec:proposal}

This section describes the \mbox{ELECTRON} system for supporting the access control in heterogeneous environments of IoT networks through devices' social relationships to enhance and simplify their recognition. \mbox{ELECTRON} aims at operating as a middleware service cooperating on the dissemination and sharing of data against impersonation threats, such as Sybil attacks. Next we detail the IoT network model where \mbox{ELECTRON} plays and describe its architecture, as well as the action behavior of Sybil attackers.

\subsection{Network model}

We assume a hybrid IoT network topology (i.e., fix and non-fix environments) in which all object devices (nodes) \mbox{collaborate} on the data forwarding service. This network supposes the existence of heterogeneous nodes (i.e., nodes with ($N_{sub}$) and without constrained resources ($N_{man}$)). These nodes exhibit random mobility in reason of ones roam to various places \mbox{corresponding} to a range of social environments. The set of IoT objects in a given instant $t$ is denoted by $\mathcal{N}$~=~$\{ n_1, n_2, ....., n_n\}$, while the set of all $N_{man}$ nodes is denoted by $\mathcal{K}$~=~$\{ N_{{man}_1}, N_{{man}_2}, ....., N_{{man}_i}\}$, and the set of all $N_{sub}$ nodes is denoted by $\mathcal{L}$~=~$\{ N_{{sub}_1}, N_{{sub}_2}, ....., N_{{sub}_j}\}$. Hence, $n_n \in  \mathcal{K}$ or $n_n \in  \mathcal{L}$, $\mathcal{K} \subseteq \mathcal{N}$ and $\mathcal{L} \subseteq \mathcal{N}$, so $\mathcal{K} \cup \mathcal{L} = \mathcal{N}$.

\begin{figure}[htbp]
 \centering
 \includegraphics[width=0.5\textwidth]{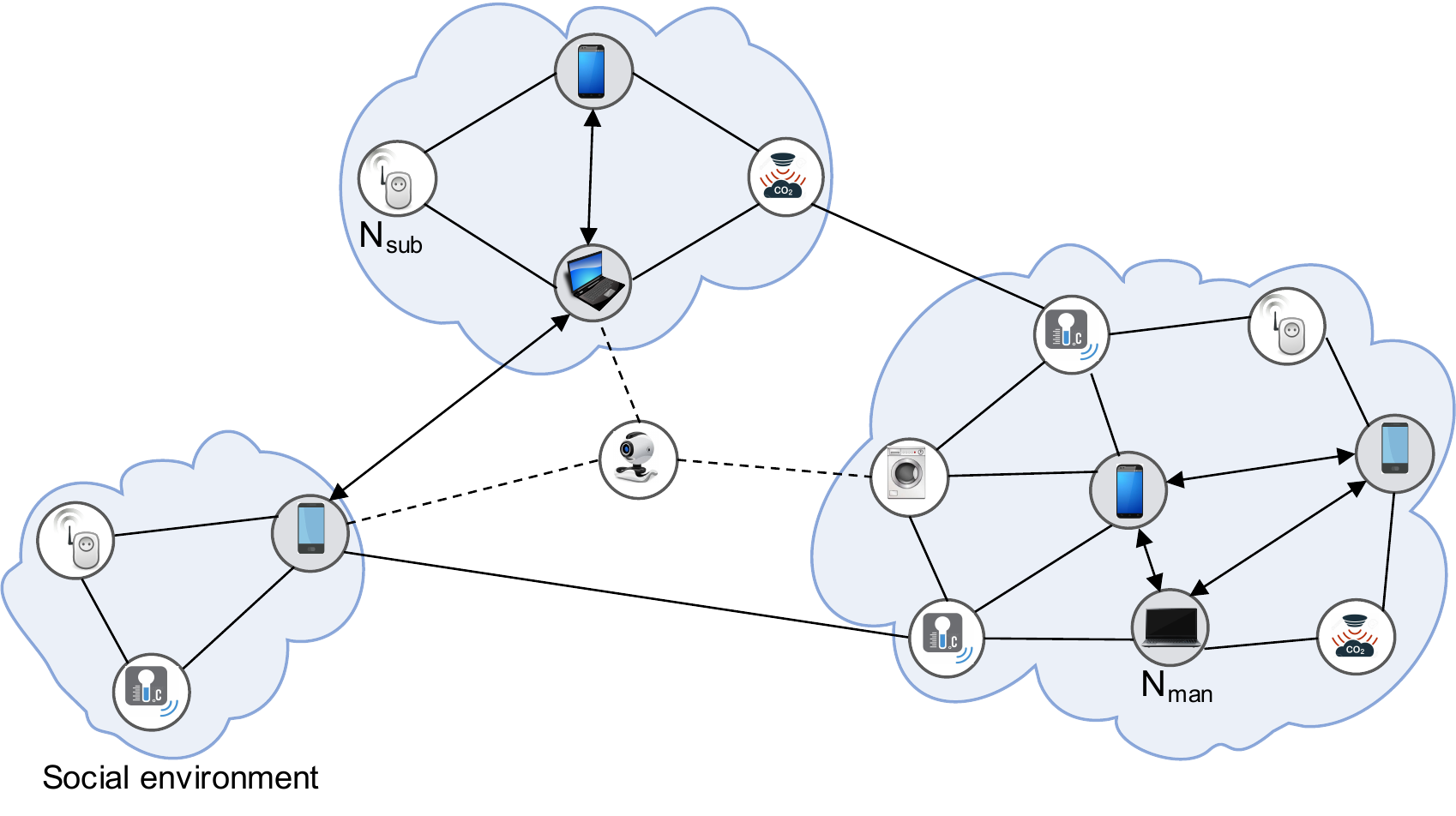}
 \caption{IoT network organization model}
 \label{fig:networkModel}
\end{figure}

As shows Figure~\ref{fig:networkModel}, $N_{man}$ nodes establish a distributed logic network to exchange and keep information about the trust and social relations management, as they own more computational resources like processing, storage, and power. These nodes control the access to the network by the information collected through all the nodes with which they interact over time. They exchange data with each other about trust and social information, which is indicated by the arrowed solid lines. We observe in the figure three distinct social environments controlled by $N_{man}$ nodes, which are established by particular social relations like friendship and common interests. The nodes with constrained resources ($N_{sub}$) are subordinate to $N_{man}$ nodes and are responsible for collecting data and sending to $N_{man}$ nodes, that forward them. While moving, $N_{sub}$ nodes request access to others contexts, as indicated by the dashed lines. Therefore, \mbox{ELECTRON} faces network scalability through a dynamic topology, in which the devices interact in specific contexts and are grouped by social relations. Further, the energy consumption of devices depends on their interactions with each other, increasing in devices that maintain strong social relationships. However, devices in different contexts or with low social interaction will exchange fewer messages, contributing to reducing their energy consumption. Further, this network organization allows to boost the coordination ability by adding a new object, $N_{man}$ node, to the management function.

\begin{figure}[!htbp]
 \centering
 \includegraphics[width=0.6\textwidth]{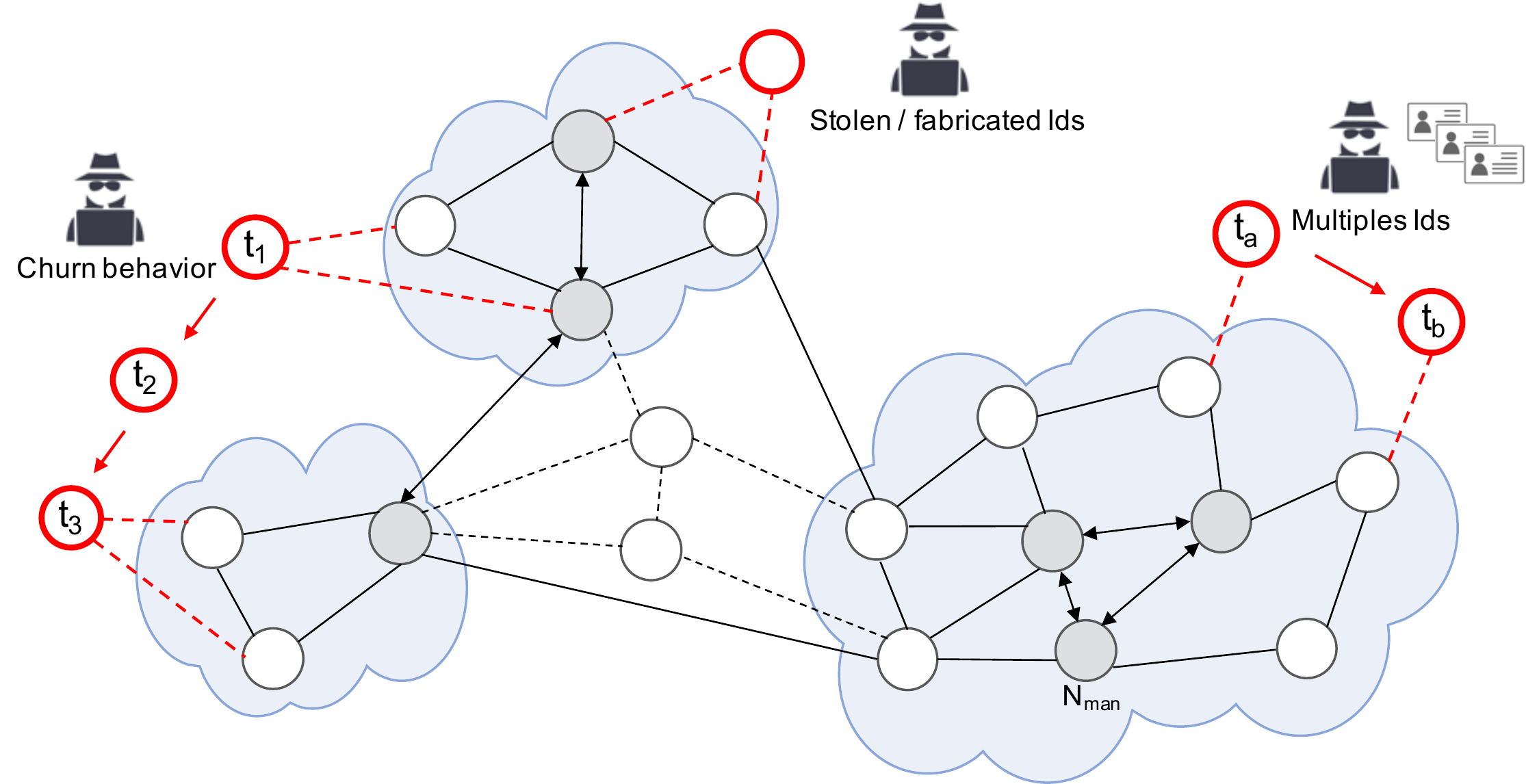}
 \caption{The behavior of the Sybil attacks}
 \label{fig:Attack}
\end{figure}

Figure~\ref{fig:Attack} presents the Sybil attackers behaviors. The attackers get access to the network with stolen or fabricated identities, in an arbitrarily and/or premeditated way. We argue that they steal identities from other legitimate nodes, while the fabricated identities do not have any previous association with the network devices. On the {\it churn} behavior, an attacker applies only one identity, and tries to establish several associations with different network nodes in a short time period. Contrarily, on the multiple identities behavior, the attacker has several identities and slower mobility. So, it tries to associate to legitimate nodes using all those identities. Moreover, an attacker can change its behavioral patterns from {\it churn} to multiple identities.

\subsection{Architecture}

\begin{figure}
 \centering
 \includegraphics[width=0.8\textwidth]{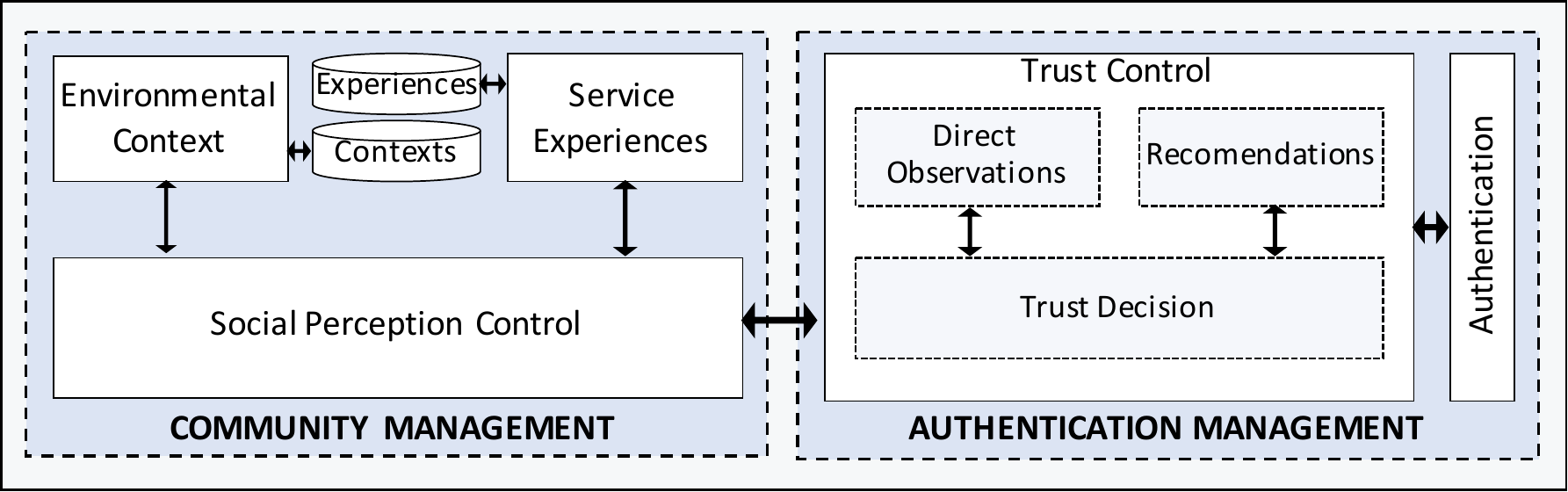}
  \caption{The \mbox{ELECTRON} architecture}
 \label{fig:Arq}
\end{figure}

\mbox{ELECTRON} comprises two modules called {\bf Community Management} (CM) and {\bf Authentication Management} (AM), as shows Figure~\ref{fig:Arq}. The CM module monitors nodes social attributes and evaluates their similarity to belong to a  community established over time based on devices interactions. The AM module coordinates the evaluation of the neighbors nodes trust and authenticates them, providing the access control to the established network. 

Those modules fully carry out into the $N_{man}$ devices, and only parts of them carry out into $N_{sub}$ due to their constrained resources. In this way, $N_{sub}$ nodes run the {\it Social Perception Control} and {\it Trust Control} components in order to forward information about social relations and experiences with others nodes in their neighborhoods to the closer $N_{man}$ device. That composition enables $N_{sub}$ to accomplish only simple tasks, while $N_{man}$ devices perform complex tasks, like trust computation and social relationships to employ such information to control the access of other nodes to the network, saving $N_{sub}$ devices resources. Next, we describe the details of each module.

\subsubsection{Community Management Module}

The CM module addresses the management of device relationships, community information, and similarities, and it comprises three components: {\it Social Perception Control}, {\it Environmental Context} and {\it Service Experiences}. 

\subsubsection*{a) Social Perception Control component}

The {\it Social Perception Control} stores data about the friendship relation and the common interests of $N_{man}$ nodes, and the relations of $N_{sub}$ nodes. Friendship is related to the number of interactions between two devices or to the relation between devices owners, as well as the interests defined by the user or extracted from his/her behavior. Such information enables the trust assessment by the {\it Authentication Management} module and contributes for verifying the similarities of nodes inside communities. These smart communities reveal the current context of the devices, playing an essential role in the social behavior of objects. 

\begin{figure}[!ht]
 \centering
 \includegraphics[width=.5\textwidth]{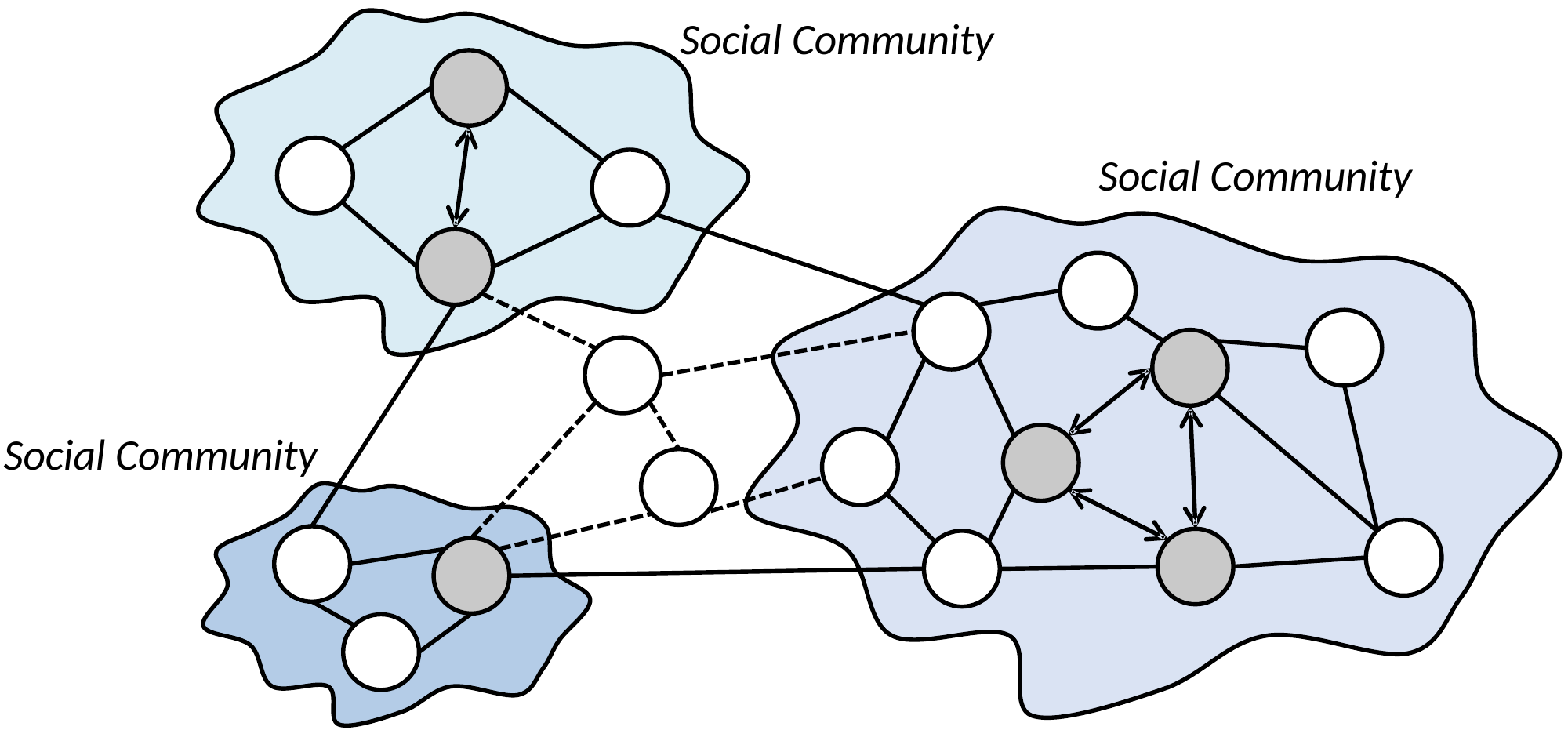}
 \caption{Communities in an SIoT network}
 \label{fig:Community}
\end{figure}

The relation among devices is stronger inside the communities, increasing the trust level with each other, mainly due to their social relation (i.e., friendship and common interests). The Figure~\ref{fig:Community} depicts a SIoT network with three distinct social communities, and $N_{man}$ (gray) and $N_{sub}$ (white) nodes. Some nodes have already established connections (solid line), as other ones are trying to get access to the network (dashed lines).

\vspace{-0.2cm}

\vspace{-0.2cm}

We describe the communities through a similarity graph $G$, where $V(G)$ is its set of vertices, and $E(G)$ is its set of edges. The edges indicate a high similarity between two vertices, as noted in Equations (\ref{eq:Similarity}) and (\ref{eq:Community}). $S(i,j)$ represents the similarity between {\it i} and {\it j} vertices of the community $\mathcal{C}$, so that values higher than the threshold ($Similarity_{threshold}$) indicate a strong similarity between the vertices, implying the creation of a community. 

\begin{equation}
  S(i,j) = Sim^\mathcal{F}(i,j) * \varphi_\mathcal{F} + Sim^{\mathcal{I}}(i,j) * \varphi_{\mathcal{I}}
  \label{eq:Similarity}
\end{equation}

\begin{equation}
  \mathcal{C} = \forall i,j \in V(G) \ | \ S(i,j) > Similarity_{threshold}
  \label{eq:Community}
\end{equation}

\vspace{0.2cm}

The tuple $\langle Sim^\mathcal{F},Sim^{\mathcal{I}}\rangle$ expresses the similarity of devices social information. The set of friendship of a device $x$ is denoted by $\mathcal{F}_x$, while the friendship similarity between two devices is represented by $Sim^\mathcal{F}$. The set of interests of a device $x$ is denoted by $\mathcal{I}_x$ and the similarity of interests between two devices is represented by $Sim^{\mathcal{I}}$. $\varphi$ denotes the weight attributed to each set, where $\varphi_\mathcal{F} + \varphi_{\mathcal{I}} = 1$. Those metrics are based on Jaccard's similarity coefficient~\cite{Abderrahim2017}, where $\mathcal{F}$ expresses a social relation that impacts on the recommendations. Social relation is tied to intimacy between entities, as in Equation (\ref{eq:friend}), where the sets $\mathcal{F}_i$ and $\mathcal{F}_j$ represent the list of friends of devices $i$ and $j$, respectively.  We obtain the  friendship similarity of these nodes ($Sim^{\mathcal{F}}(i,j)$) by
Equation~(\ref{eq:friend}).

\begin{equation}
    Sim^\mathcal{F}(i,j) = \frac{|\mathcal{F}_i \cap \mathcal{F}_j|}{|\mathcal{F}_i \cup \mathcal{F}_j|}
    \label{eq:friend}  
\end{equation}

Inside a community, nodes share common interests, which increases the probability to know habitual patterns about services provided by a device. We obtain the interests similarity of nodes {\it i} and {\it j} ($Sim^{\mathcal{I}}(i,j)$) by Equation~(\ref{eq:interest}), from their individual shared interests, $\mathcal{I}_i$ and $\mathcal{I}_j$, respectively.

\vspace{0.2cm}

\begin{equation}
    Sim^{\mathcal{I}}(i,j) = \frac{|\mathcal{I}_i \cap \mathcal{I}_j|}{|\mathcal{I}_i \cup \mathcal{I}_j|} 
    \label{eq:interest}
\end{equation}

\vspace{0.2cm}

SIoT relations~\cite{Atzori2011,atzori2012social,nitti2013trustworthiness} can be classified in {\it parental object relationship} (POR), {\it co-location object relationship} (CLOR), {\it co-work object relationship} (CWOR), {\it ownership object relationship} (OOR), and {\it social object relationship} (SOR). As similar objects establish homogeneous relations, devices built by a manufacturer in the same period form {\it parental relationships} (POR). This kind of relationship is already implemented during production phase, and it will not change over time, but updated by events like disruption/obsolescence of a given device. The {\it co-location relationships} (CLOR) happen among objects that are always at the same place (e.g., sensors, actuators, refrigerators, and televisions), as in smart homes and smart cities. The {\it co-work relationships} (CWOR) always start when objects collaborate to provide a common application in IoT. {\it Ownership relationships} (OOR) happen with objects from the same owner, commonly among heterogeneous objects (i.e., smartphones, computers, and game consoles). Lastly, the {\it social relationships} (SOR) are tied to their object owners' relations and arise when the objects interact with each other. Thus, they build this relation when come into contact, sporadically or continuously, for purposes solely related to relations among their owners (e.g., devices/sensors belonging to friends).

\subsubsection*{b) Environmental Context and Service Experiences components}

The {\it Environmental Context} (EC) component identifies and informs about the current context of the device, which represents a vision based on information like current device location and near devices. It also brings detailed information about locations devices have passed by, as the current context of the device influences the trust computation and its convergence. Thus, knowing the context and adapting to it have been the subject of many works.~\cite{Fernandez2017} However, natural dynamics in the environment of IoT networks, mainly due to objects mobility, requires dealing with several different contexts. Hence, it becomes necessary to receive the information and to identify the actual context to improve the accuracy of the evaluation.

EC component deals with a detailed list about the environments in which the device is located, such as places and devices types. Thus, this component collaborate to the social trust evaluation, bringing the context influence to this process, increasing the relevance of the environments in which devices operate. It offers the base value $a$ to the trust evaluation, through the analyses of the context of the communities devices. According to this base value, it improves or worsens the trust convergence.

The {\it Service Experiences} component stores bad experiences on services provided by neighbors devices, which means maintaining a history of interactions with devices that have provided a poor quality of service. This strategy is based on a psychological approach, in which people tend to keep the peculiarities of the environment that caused negative interactions.~\cite{mcknight1998} Hence, this component is strongly tied to the context, since bad experiences keep relation to the current context. In this way, in future interactions, it is possible to have better knowledge to assign lower trust values in similar conditions by the association of the bad experiences with the context.

\subsubsection{Authentication Management Module}\label{sec:AMM}

This module manages the devices authentication for controlling their access to the network by the {\it Trust Control} and {\it Authentication} components. The {\it Trust Control} component assesses the current devices' trust to support the decision-making in controlling their access to the network.  It comprises the {\it Direct Observations}, {\it Recommendations} and {\it Trust Decision} subcomponents, so that the first one monitors support the direct trust evaluation, while the second one controls the opinions about other devices received from neighbors devices to support trust evaluation. The last one evaluates devices trust by taking into account the information received from other subcomponents and from the {\it Community Management} module, and send it to the {\it Authentication} component.

The use of trust in the IoT context is not entirely new, but there is a strong tie between trust and the social context of the objects in this work. We aim that the sociability and the relationship established over the time support to gather more precise values about devices' behavior. We have adapted the subjective logic~\cite{Josang2016book} to incorporate the influence of similarity and relationship to trust assessment, as in Equation~(\ref{eq:OverallTrust}). The general trust of node $i$  about node $j$ ($T_{ij}^{\mathcal{C}}$) takes into account the {\it Direct Trust} ($D_{ij}$) and the {\it Recommendation} ($R_{ij}$). {\it Direct Trust} ($D_{ij}$) deals with the experiences of devices and its neighbors, and brings the interactions between devices $i$ and $j$, as in Equation~(\ref{eq:DirectTrust}). This value is essential in the trust assessment because it demonstrates the outcomes of device choices. {\it Community Similarity} ($S_{ij}^{\mathcal{C}}$) indicates how much a device $j$ resembles other devices belonging to the same community $\mathcal{C}$. This property influences the current trust through factors like common interest and friendship, and improves its accuracy, as in Equation~(\ref{eq:Similarity}). Recommendation ($R_{ij}$) is the indirect trust of device $i$ about device $j$, which means that device $i$ builds its trust about device $j$ taking into account the shared opinions from other communities devices about device $j$. We obtain $R_{ij}$ by Equation~(\ref{eq:Recommendation}).

\begin{equation}
    \centering
 	T_{ij}^{\mathcal{C}} =  \alpha D_{ij} + \beta S_{ij}^{\mathcal{C}} + \gamma R_{ij}
\label{eq:OverallTrust}
\end{equation}

\begin{equation}
    \centering
 	D_{ij} =  b + d + u
 	\label{eq:DirectTrust}
\end{equation}

\begin{equation}
    \centering
 	R_{ij} = b + d + u     
    \label{eq:Recommendation}
\end{equation}

\vspace{0.3cm}

{\it Direct trust}, {\it Similarity}, and {\it Recommendation} influence general trust in a different way. While recommendations rely on the social relation between the recommender and the requester, direct trust and similarity are tied mainly to the requester. So, both of them have a more significant impact on general trust.  We have assigned in this work values for each kind of relation, as shown in Table~\ref{tab:RelacaoSIoT},~\cite{Abderrahim2017} representing a relationship factor in trust composition, as seen in Equation~(\ref{eq:OverallTrust}). A relationship factor ($\gamma$) weights the recommendations to filter the most important ones to trust through social relations, as $\alpha$ and $\beta$ weights direct trust and similarity, respectively. As the sum of weights is equal to 1 ($\alpha + \beta + \gamma = 1$), $\alpha$ and $\beta$ receive the same values, corresponding to the division of the excess weight to the $\gamma$ value assigned in Table~\ref{tab:RelacaoSIoT}.

\begin{table}[htbp]
  \centering
  {\scriptsize
   \caption{Relationship Factors}
   \label{tab:RelacaoSIoT}
  \begin{tabular}{lccccc} \hline
  & \multicolumn{5}{c}{{\bf Objects relationship}}  \bigstrut \\ \cline{2-6}
    & CLOR  & CWOR & OOR & SOR & POR \bigstrut \\ \hline 
   {\bf Relationship factor ($\gamma$)}      & 0.3   & 0.2 & 0.2 & 0.1 & 0.1 \bigstrut \\ \hline
  \end{tabular}}
\end{table}

\begin{figure}[htbp]
 \centering
 \includegraphics[width=.4\textwidth]{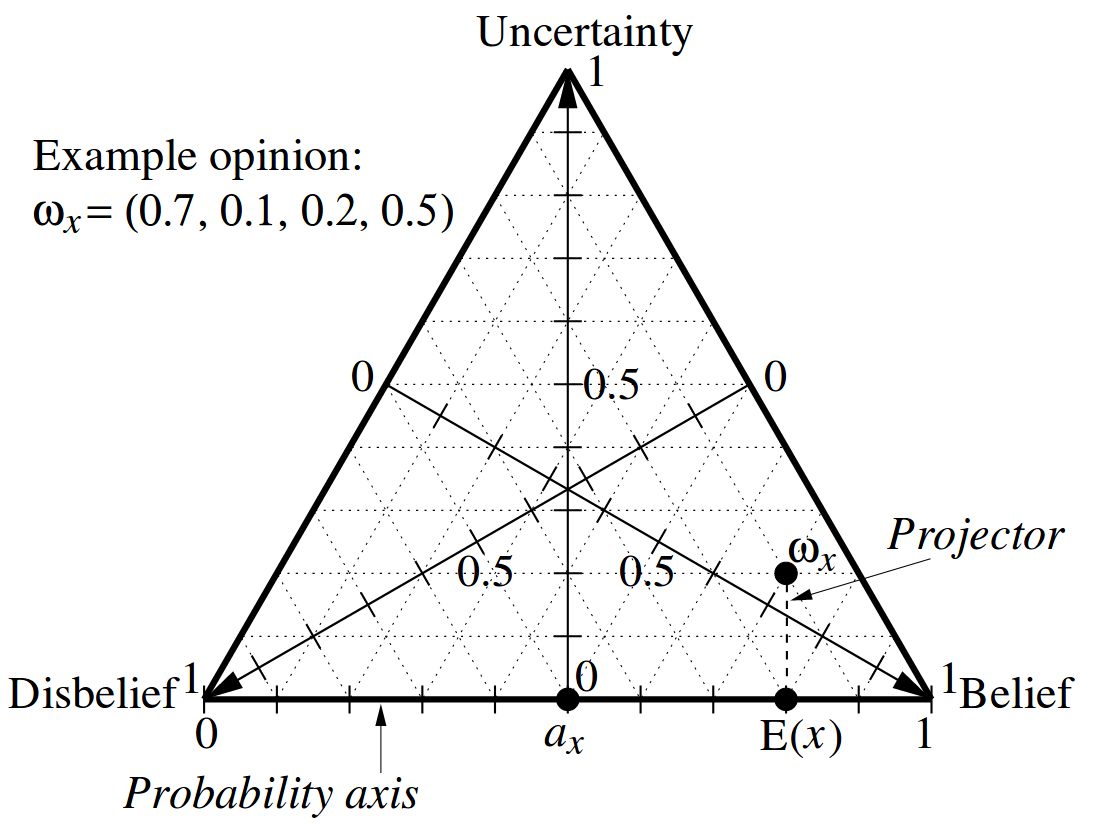}
 \caption{Triangle of opinions~\cite{Josang97}}
 \label{fig:OpinionTriangle}
\end{figure}

The {\it Subjective Logic} (SL) has been applied in IoT to manage trust through opinions about other network devices.~\cite{Son2017} In this way, as shows Figure~\ref{fig:OpinionTriangle}, we establish a triangle of opinions~\cite{Josang97} with the variables belief ({\it b}), unbelief ({\it d}), and uncertainty ({\it u}). We obtain the values of $b,\;d,\;u$ by Equations~(\ref{eq:Opnionb}-\ref{eq:Opnionu}), where $b+d+u=1$, and $\{b,d,u\} \in [0,1]^3$.~\cite{Josang2016book} The opinion at the SL computes positive and negative experiences about a given device. We have applied a SL model that encompass a 4-tuple $\langle b,d,u,a \rangle$, in which $(b+d+u)=1$, $a=[0,1]$, and the trust about an opinion is expressed by the expected value $b+u*a$. The $pos$ and $neg$ values express the amount of positive and negative experiences, respectively. The {\it Environmental Context} component defines the $a$ value, which represents the base rate to the trust convergence. The $a$ values closer to 1 represent more secure contexts (e.g., residences), while the values closer to 0, the less secure ones (i.e., open environments like parks).

\begin{equation}
  b = \frac{pos}{(pos+neg+2.0)}
  \label{eq:Opnionb}
\end{equation}

\vspace{0.3cm}

\begin{equation}
  d = \frac{neg}{(pos+neg+2.0)}
  \label{eq:Opniond}
\end{equation}

\vspace{0.3cm}

\begin{equation}
 u = \frac{2.0}{(pos+neg+2.0)}
  \label{eq:Opnionu}
\end{equation}

\vspace{0.3cm}

The {\it Authentication} component is responsible for the decision making on the access control in \mbox{ELECTRON} based on the trust value sent by the {\it Trust Control} component. We have set up that trust values above $0.6$ allow us access to all services and applications available on the network. Though, it is possible to add other techniques to this component like fuzzy logic, for instance, to support the decision making.

\subsection{Operation}

\begin{figure}[!ht]
 \centering
 \includegraphics[width=0.9\textwidth]{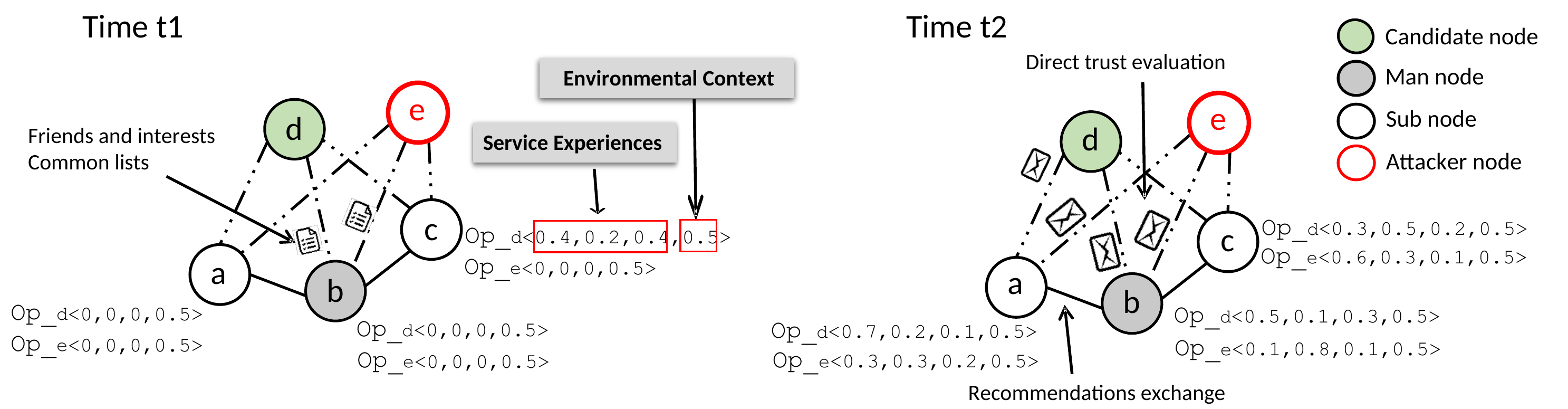}
 \caption{Creating opinions about candidates nodes}
 \label{fig:operation1}
\end{figure}

We exemplify the \mbox{ELECTRON} operation by considering an illustrative scenario in a school environment where five people (nodes) - $N_a, N_b, N_c, N_d$ and $N_e$ - interact to each other in a community of interest with a base rate ($a$) of 0.5, as depicts Figure~\ref{fig:operation1}. $N_d$ is a candidate node aiming to participate of the network, while $N_e$ is a Sybil attacker. Initially, in time $t1$, the candidate node ($N_d$) send to network nodes its lists of friendship and common interests, which means teachers and students, and subjects, respectively. Then, the {\it Trust Control} component in the $N_{man}$ node $N_b$ verifies the opinion of others network nodes about $N_d$. The opinions start with value 0 and suffer the influence of the environment context and service experiences, as the context defines the $a$ value according to the community a node request the access and its history of interactions. This history allows to compute the previous positive and negative experiences to the new social trust about this device. An opinion is a tuple $\langle$\textit{experiences,context}$\rangle$ describing all the previous experiences of a node about another one and the base rate for their context, i.e, the base rate $a$. For example, as shows Figure~\ref{fig:operation1}, the $N_c$ opinion about $N_d$ ($Op\_d$) is $\langle$0.4,0.2,0.4,0.5$\rangle$, indicating that have already interacted three times in this context.

The interactions between network nodes and $N_d$ start right after its trust evaluation. As Figure~\ref{fig:operation1} shows in time $t2$, network nodes begin evaluating the direct trust about $N_d$, computing positives and negatives experiences over time as long as they interact to each other. From these experiences, each network node calculates the direct trust about $N_d$ by Equations~\ref{eq:Opnionb},~\ref{eq:Opniond} and \ref{eq:Opnionu}. Later, network nodes $N_a, N_b$ and $N_c$ exchange their recommendations based on the calculated direct trust. With this information, each one of them calculates the first value for the social trust to $N_d$. This value depends on the social relation established with the neighbor node, besides the similarity between $N_d$ and the community, as denotes Equation~\ref{eq:OverallTrust}. Their interactions over time change their opinions about each other, as shows Figure~\ref{fig:operation1} in time $t2$. The $N_c$ opinion about $N_d$ change to $\langle$0.3,0.5,0.2,0.5$\rangle$.

\begin{figure}[!ht]
 \centering
 \includegraphics[width=0.7\textwidth]{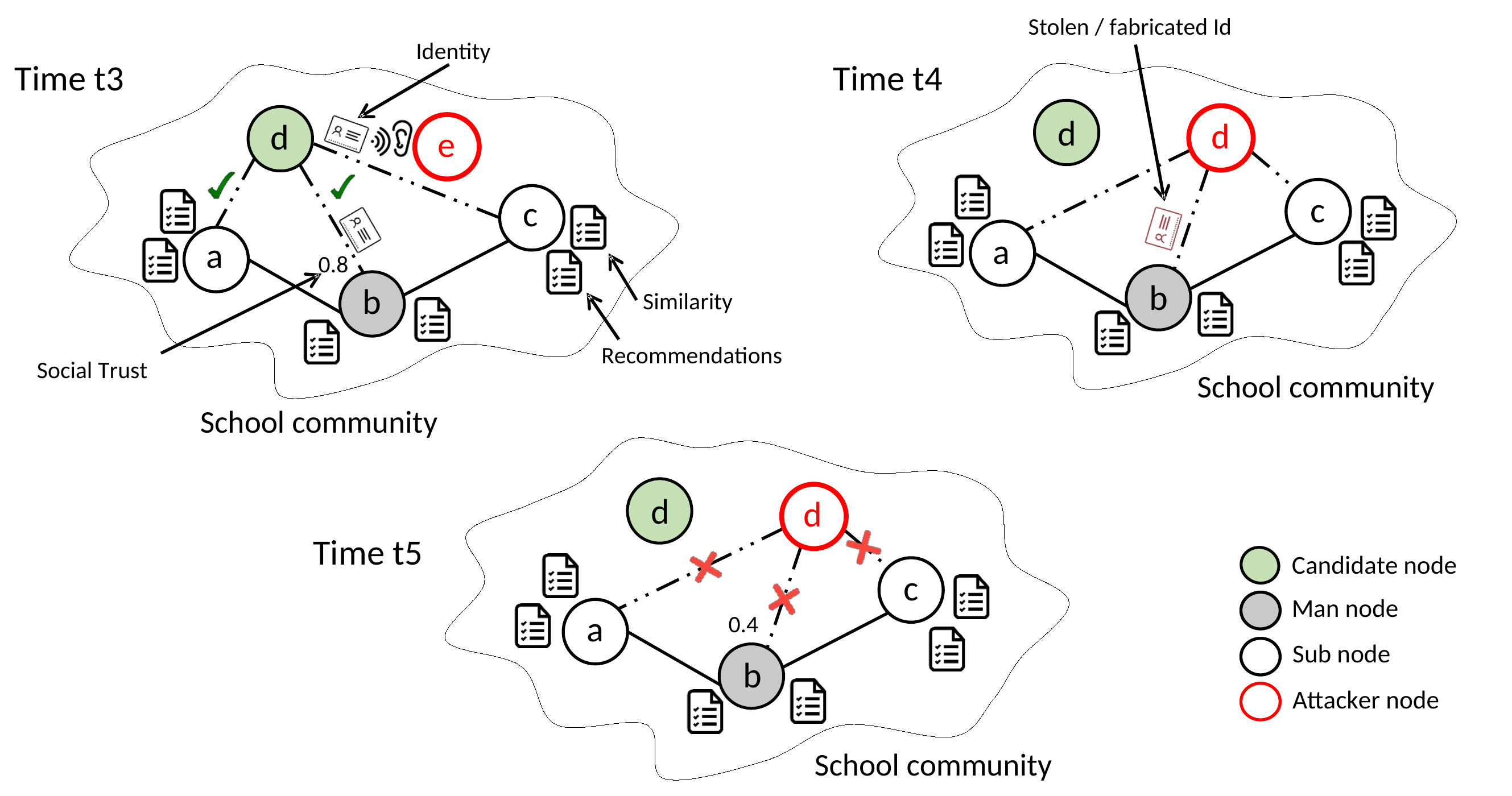}
 \caption{Detection of Sybil attack with stolen/fabricated identities}
 \label{fig:operation2}
\end{figure}

Node $N_e$ is a Sybil attacker which aims to steal identities by a promiscuous operation, as Figure~\ref{fig:operation2} shows in time $t3$. It fabricates new identities from the stored stolen identities. Next, in time $t4$, $N_e$ tries to authenticate itself in the network requesting access by a stolen identity. However, the {\it Authentication Management} in $N_b$ verifies that $N_e$ trust did not achieved a minimum value of 0.6. Hence, as Figure~\ref{fig:operation2} shows in $t5$, based on $N_e$ behavior, its social relations and similarity with the community, \mbox{ELECTRON} blocks $N_e$ access to the network, preserving its privacy.

\section{Performance Evaluation}\label{sec:eval}

This section presents an evaluation of the \mbox{ELECTRON} performance against Sybil attackers on the IoT network, and a performance comparison with \mbox{SA$^{2}$CI} system.~\cite{Evangelista2016IEEE} We initially describe the simulation scenarios and the metrics employed, and next present and discuss the achieved results.

\subsection{Simulation Settings}

We have implemented the \mbox{ELECTRON} mechanism and Sybil attackers' behavior in the NS-3 simulator,~\cite{NS3simulator2018} version 3.27, as well as we carry out the simulations of IoT scenarios. Our analysis takes into account the behavior on five types of social communities - {\it residence}, {\it school}, {\it office}, {\it gym}, and {\it park} - and five distinct social relations in SIoT - {\it parental} (POR), {\it co-location} (CLOR), {\it co-work} (CWOR), {\it ownership} (OOR), and {\it social} (SOR). We also carried out a comparative analysis between \mbox{ELECTRON} and \mbox{SA$^{2}$CI}, since \mbox{SA$^{2}$CI} also prevents Sybil attacks in IoT, but it employs distinct techniques to identify attackers like Elliptic Curve Cryptography (ECC) and Physical Unclonable Functions (PUF). Over their operation the Sybil attackers employ stolen and fabricated keys to get access to the network and exhibit {\it churn} and multiple identities behaviors in order to mask their actions. The scenario relies on a dataset from Brightkite social network,~\cite{stanfordDataset2011} in which users share their location with friends through check-ins in areas like San Francisco City (USA). The dataset encompasses environments like residences, offices, leisure places, among others. The friendship network gathered information~from 2008 to 2010 and contains 58~228 nodes and 214~078 edges, where each edge means a friendship tie. Thus, we have associated friendship information to the IoT devices in the simulation aiming to mimic a realistic social~environment.

We have also assigned security levels to the social communities environments. As detailed in Subsection~\ref{sec:AMM}, the tuple  $\langle b,d,u,a \rangle$ means information about the neighbor opinion, so that $a$ value corresponds to a value of an empiric security level inside the interval [1,0]. The values start from 1 for higher secure environments, in which devices are well known (i.e., domestic environment), and decreases to 0 in lowest secure ones, where there are unfamiliar devices (i.e., outdoor environments).

The devices make use of the IEEE 802.15.4 standard in a 6LowPAN network and employ UDP transport protocol. Thus, they disseminate information resembling a real-time application in an IoT environment. The evaluated scenarios encompass an area of 100 m x 100 m, where all nodes move with 2 m/s speed up in a {\it Random Waypoint} mobility model. We have conducted simulations in NS-3~\cite{NS3simulator2018} with 100, 150, and 200 nodes and with $a$ values of  1, 0.7, 0.5, 0.4 e 0.2, meaning {\it residence}, {\it office}, {\it school}, {\it gym}, and {\it park}, respectively. In each simulation round, whose time is 600 seconds, 10\% of nodes act as attackers and try to access the network.

\subsubsection{Metrics}

We have took into account the social issue in reason of the users mobility and its ability to protect the environment from Sybil attacks, i.e., attacks that forge a true or false identity. Thus, we employ the metrics ~{\it Evolution of Social Trust} (ESR), {\it Attack Detection Rate} (DR),  {\it Accuracy}  (ACC), {\it False Positive Rate}  (FP), and  {\it False Negative Rate} (FN).  We assess all metrics in \mbox{ELECTRON} and \mbox{SA$^{2}$CI}, except ESR, it was applied only to \mbox{ELECTRON}. 

The {\it Evolution of Social Trust}  (ESR) indicates how the social trust evolves over time through a cumulative distribution function (CDF), being obtained by Equation~(\ref{eq:OverallTrust}). ESR enables us to observe the behavior of social trust in each community, since the values of the base rate of each community are unique, and thus they influence the trust evolution. We have filtered recommendations by social relationships to analyze mechanism behavior according to relations established in each one of these conditions.

The {\it Attack Detection Rate} (DR) computes the percentage of all Sybil attacks effectively identified. It is a ratio between the amount of attacks correctly detected (ACD) under the number of attacks existing over time (AOT), as in Equation~(\ref{eq:det}). 

\begin{equation}
    DR =  \frac{ACD}{AOT}\;\times\; 100\%
 \label{eq:det}
\end{equation}  

\vspace{0.3cm}

The {\it Accuracy} (ACC) indicates the mechanism precision in detecting Sybil attacks. It means the ratio between the sum of all Sybil attacks correctly detected (ACD) and the legitimate nodes correctly identified (LCI), under all requests sent to the network (AR), as shows Equation~(\ref{eq:ac}). ACC presents discrete values between 0 and 1, and the closer they are to 1, the better the accuracy.

\begin{equation}
	ACC = \frac{ACD + LCI}{AR}
	\label{eq:ac}	
\end{equation}  

\vspace{0.3cm}

The {\it False Negative Rate} (FN) computes the number of times the attackers were classified as legitimate nodes. It means a ratio between the amount of attacking nodes incorrectly classified  ({\it FalseNeg}) and the sum of the {\it FalseNeg} and the number of attacking nodes correctly classified ({\it TruePos}) (i.e., legitimate nodes correctly identified). FN is obtained by Equation~(\ref{eq:fn}). 

\begin{equation}
	FN = \frac{FalseNeg}{FalseNeg+TruePos} \times 100\%
	\label{eq:fn}
\end{equation}  

\vspace{0.3cm}

The {\it False Positive Rate} (FP) computes the number of times that Sybil attacks were incorrectly identified. It means a ratio between the number of Sybil attacks incorrectly identified (\textit{FalsePos}) and the sum of the \textit{FalsePos} and the number of Sybil attacks not succeeded and correctly identified (\textit{TrueNeg}). FP is measured by Equation~(\ref{eq:fp}).

\begin{equation}
	FP = \frac{FalsePos}{FalsePos+TrueNeg} \times 100\%
	\label{eq:fp}
\end{equation}

\vspace{0.3cm}

\subsection{Social Perception Analysis}

\begin{figure}
    \centering
    \includegraphics[width=65mm]{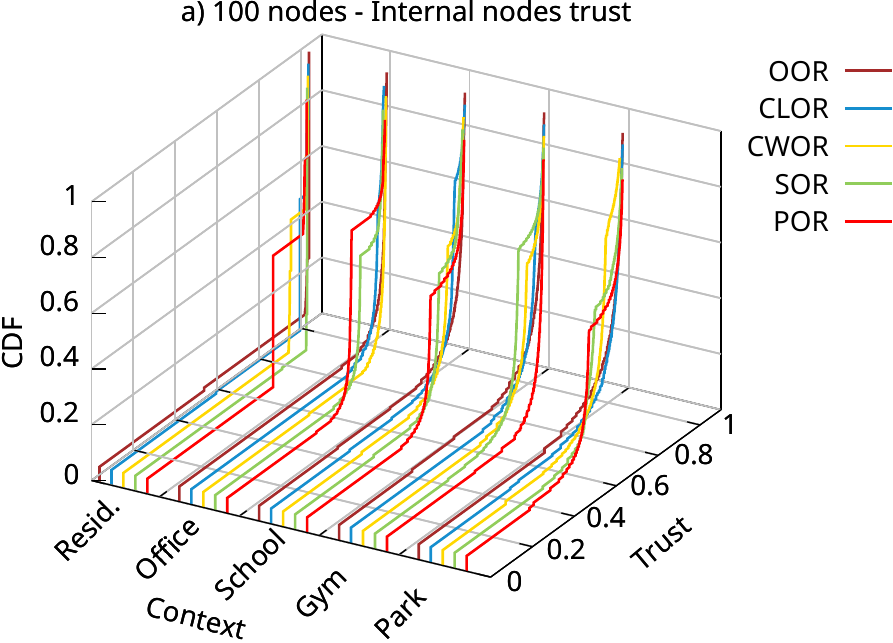}
    ~
    \includegraphics[width=65mm]{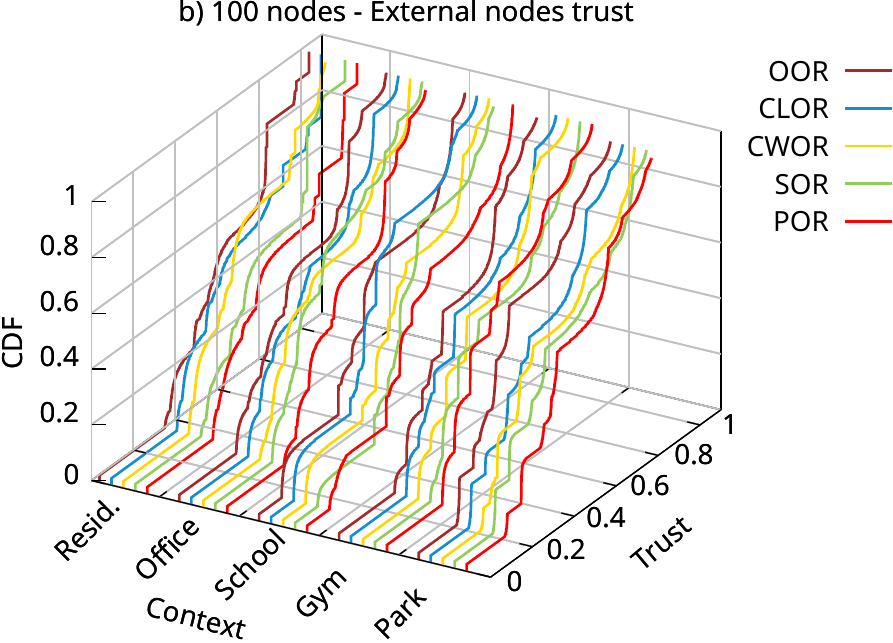}
    \vspace{0.5cm}
    
    \includegraphics[width=65mm]{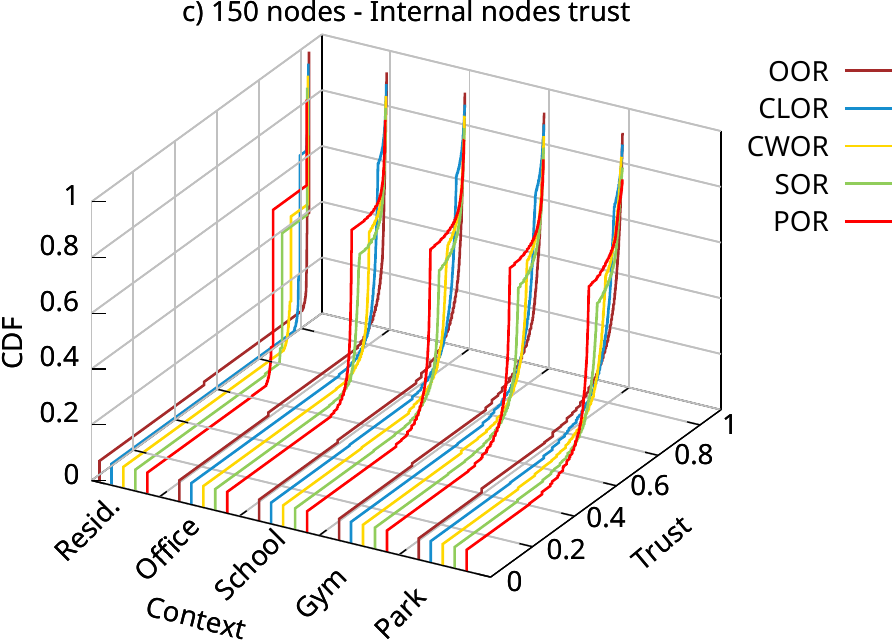}
    ~
    \includegraphics[width=65mm]{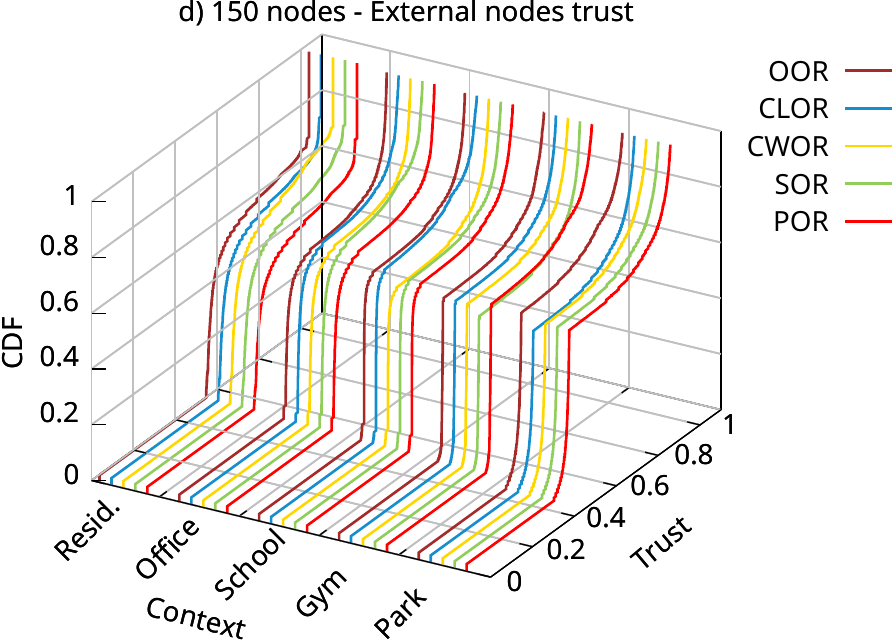}
    \vspace{0.5cm}
    
    \includegraphics[width=65mm]{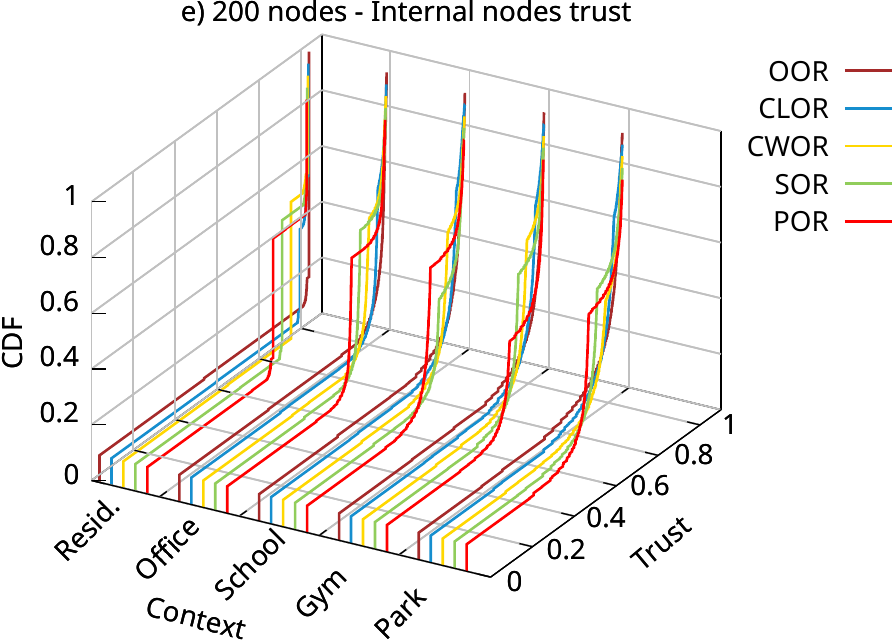}
    ~
    \includegraphics[width=65mm]{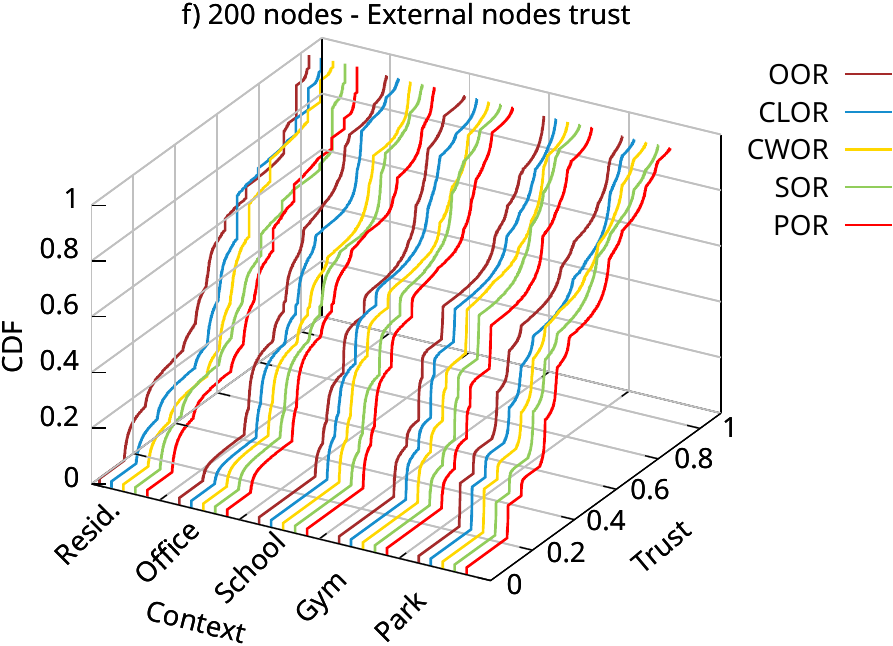}
    \caption{Social trust evolution in scenarios with 100, 150 and 200 nodes}
    \label{fig:CDF100ChurnStolen}
\end{figure}

We analyze the devices social behavior in \mbox{ELECTRON} by the social trust values obtained through Equation~(\ref{eq:OverallTrust}) in five distinct social relations in SIoT: {\it parental} (POR), {\it co-location} (CLOR), {\it co-work} (CWOR), {\it ownership} (OOR), and {\it social} (SOR). Each one of these relations affects the trust, as we filter recommendations against the social relation built with the neighbor node. The graphics depicted in Figure~\ref{fig:CDF100ChurnStolen} show the {\it Evolution of Social Trust} (ESR) of each one of the evaluated communities, taking into account their context and the base rate $a$ to obtain the trust value. We represent each community by pairs of graphics using CDFs, where the left one bring the trust evolution among nodes inside the network, and the right one focus on the outside nodes requesting access to the network. Moreover, we consider scenarios with attackers exhibiting  {\it churn} behavior and stolen identities, as these kind of attacks mean the most significant challenges to \mbox{ELECTRON}. 

Figures~\ref{fig:CDF100ChurnStolen}a and \ref{fig:CDF100ChurnStolen}b show the social trust built among network nodes and the outside ones for a scenario with 100 nodes. We observe that \mbox{ELECTRON} presents the worst performance in parental relation for the internal nodes, mainly because this relation does not give so much importance to recommendations, and have presented a better result for {\it gym} community. However, this behavior changes among outside nodes, achieving higher values for {\it residence} community and lower ones for {\it gym} community. The social trust evolution presents an unstable behavior among network nodes, as we verify by their higher trust values. This performance is due to attackers' behavior, which frequently brings down the aggregated social trust value. 

Figures~\ref{fig:CDF100ChurnStolen}c and~\ref{fig:CDF100ChurnStolen}d present the social trust evolution of the external nodes for scenarios with 150 nodes. While in scenarios with 100 nodes, the values of social trust start between 0.1 and 0.3, here they start between 0.4 and 0.5. This behavior is due to the distribution of nodes in network nodes, legitimate nodes to access the network, and attackers nodes. We set up the number of attacker nodes bigger than legitimate ones, making the trust values unstable and tending to lower values. In scenarios with 150 nodes, more legitimate nodes request access to the network, hence collaborating to increase trust values with more constancy.

Figures~\ref{fig:CDF100ChurnStolen}e and~\ref{fig:CDF100ChurnStolen}f present the social trust evolution of the external nodes for scenarios with 200 nodes. The behavior for internal nodes is similar to that with 150 nodes, but we observe a distinct trust evolution with external nodes. Through Figure~\ref{fig:CDF100ChurnStolen}f, we verify that trust starts around 0.2, but it presents an unstable behavior among network nodes, as in the scenario with 100 nodes (Figure~\ref{fig:CDF100ChurnStolen}b). This performance is due to attackers' behavior, which frequently brings down the aggregated social trust value.

\subsection{Comparative Analysis}

We conducted an extensive comparison between \mbox{ELECTRON} and \mbox{SA$^{2}$CI} to analyze their performance in face of Sybil attacks. The analysis has taken into account the following metrics: {\it Detection Rate} (DR), {\it Accuracy} (ACC), {\it False Negative Rate} (FN) and {\it False Positive Rate} (FP). We analyzed the Sybil attackers detection in scenarios with 100, 150, and 200 nodes with all behaviors facing 10\% of attackers. For the sake of space and due to the great volume of gathered data, the other metrics refer to Sybil attacks with {\it churn} behavior using stolen identities over a simulation with 100 nodes, and multiple fabricated identities over a simulation with 150 nodes. We choose these behaviors settings due to the highest challenge to \mbox{ELECTRON} and to provide an effective analysis.

\begin{figure}
    \centering
    \includegraphics[width=65mm]{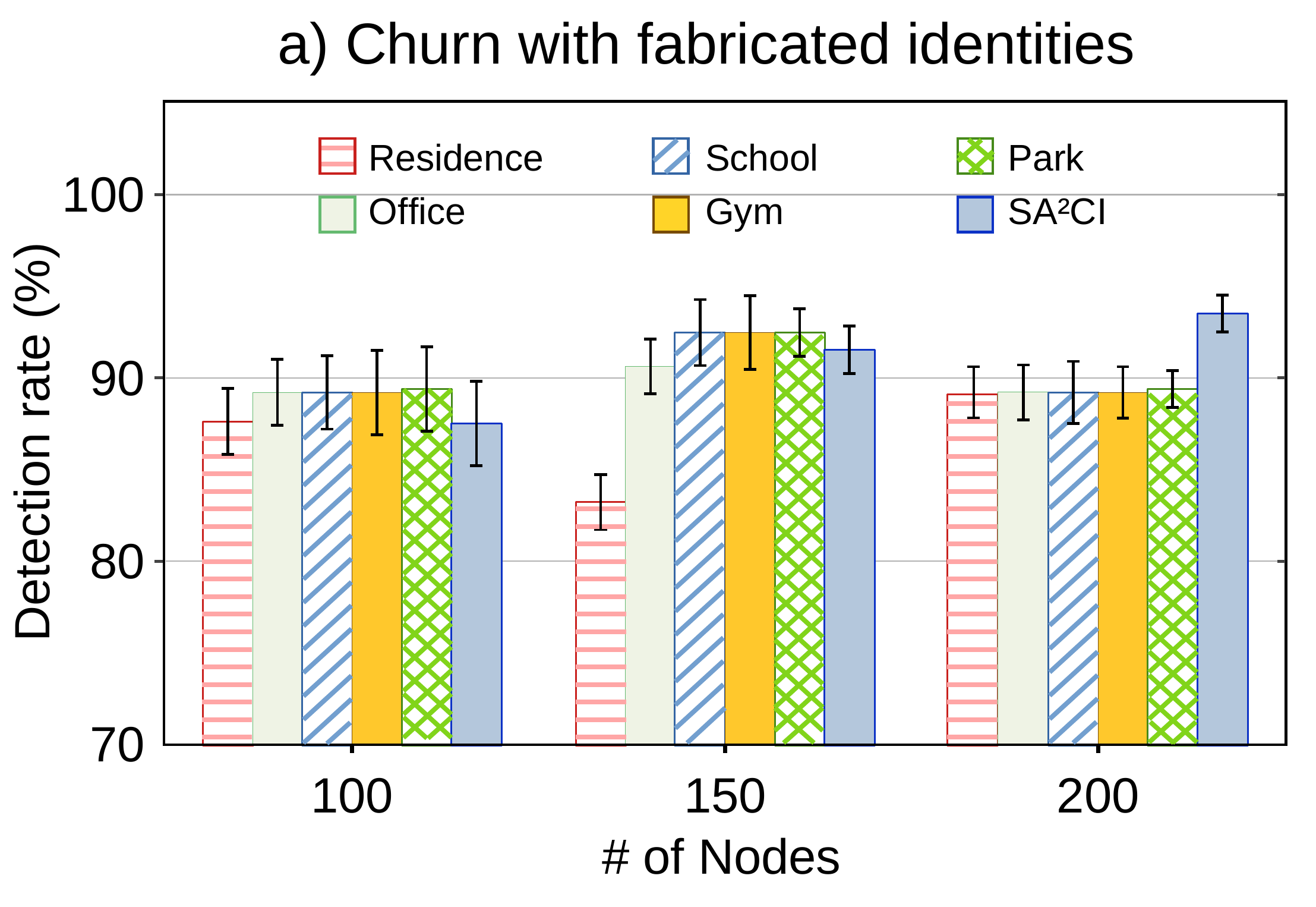}
    ~
    \includegraphics[width=65mm]{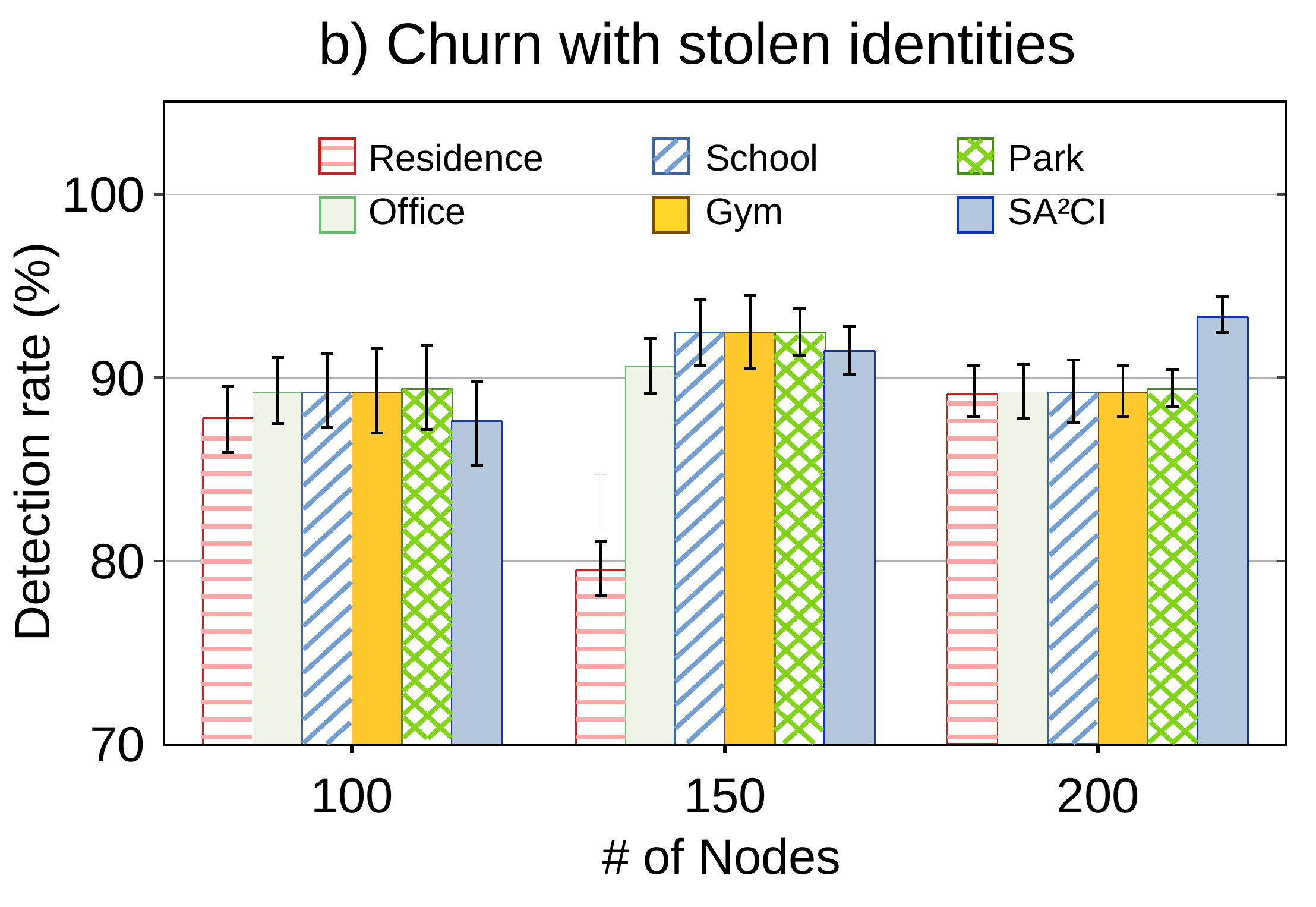}
\end{figure}

\begin{figure}
    \centering
    \includegraphics[width=65mm]{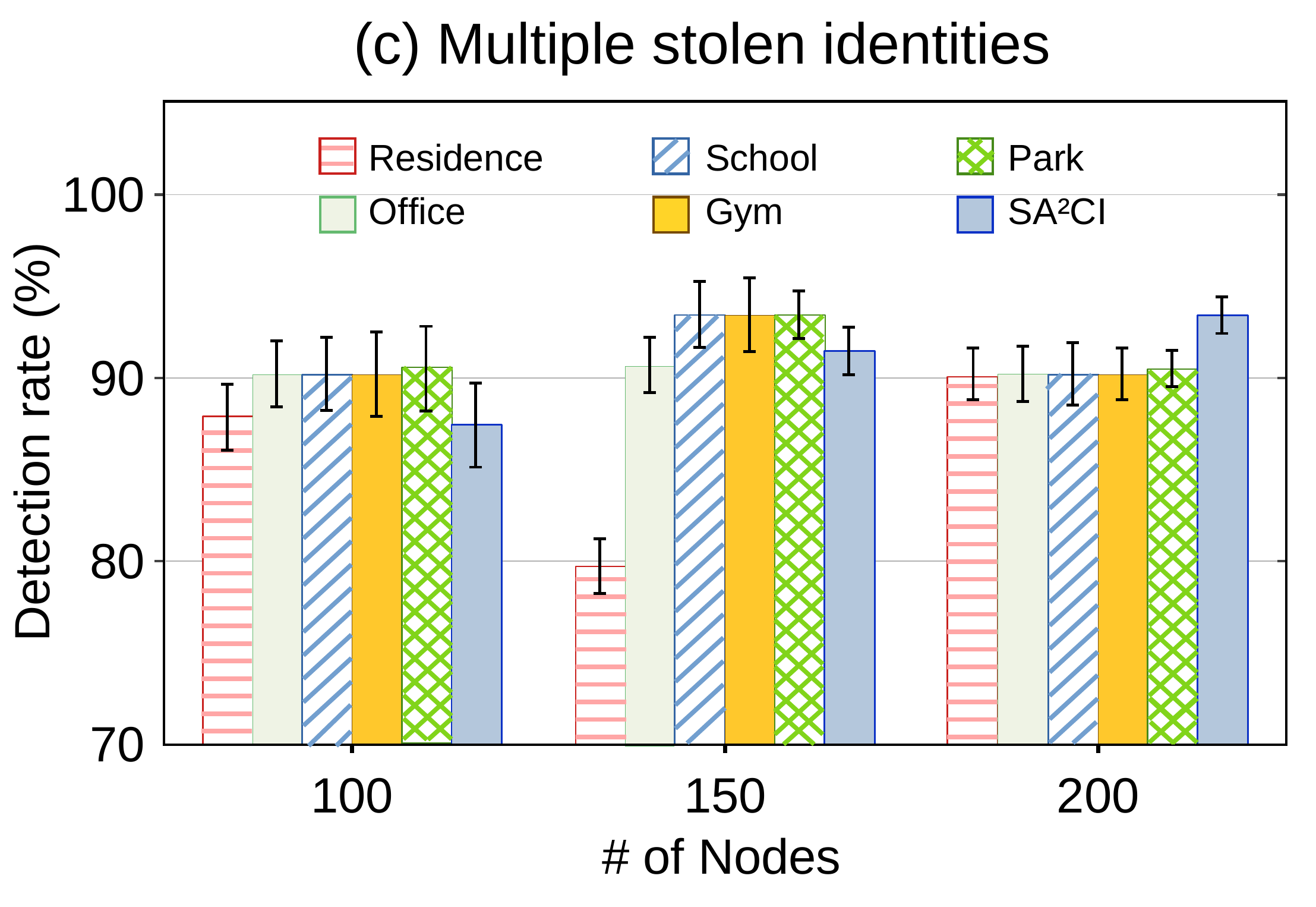}
    ~
    \includegraphics[width=65mm]{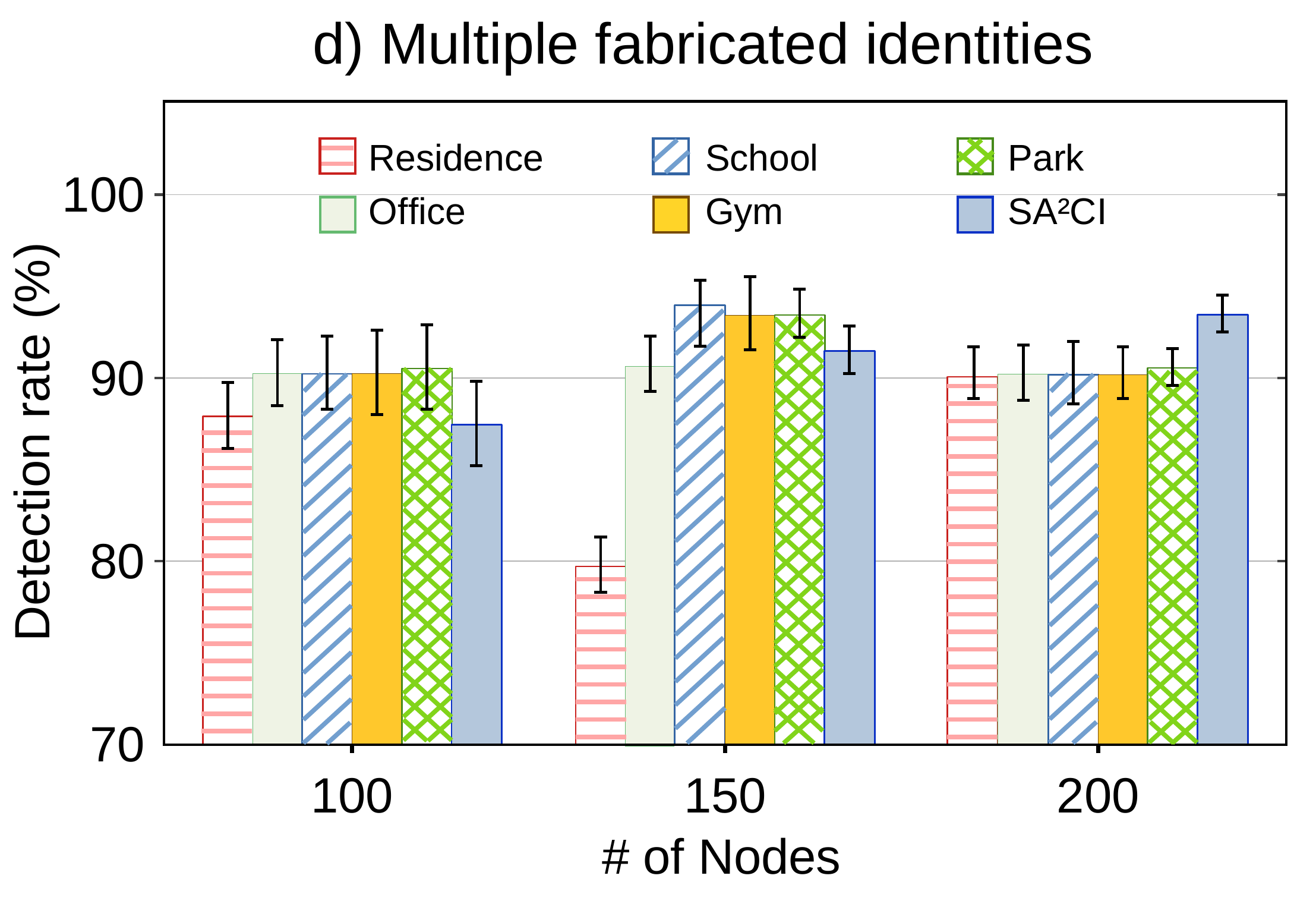}
    \caption{Attacks detection rate}
    \label{fig:detectedRate}
\end{figure}

We verify the attack detection rate for \mbox{ELECTRON} and \mbox{SA$^{2}$CI}, whose results are shown in Figure~\ref{fig:detectedRate}. We can see that they achieved DR values between 79.6\% and 92.5\%, keeping stable over the simulation. In a setting with high mobility and stolen identities, \mbox{ELECTRON} needed more time to detect a Sybil attacker because the social trust value is an active principle for identifying attackers, being expected then a gradual threat detection. The graphs show better performance of \mbox{ELECTRON} friendships and common interests in the building of the social trust. \mbox{ELECTRON} achieved a DR around 90.2\% in the scenario with 200 nodes and shows a higher stability when compared to other scenarios. This behavior is due to the trust value measured during the attacks, mainly into the {\it residence} and {\it office} communities. At the time the attacker achieved trust values higher than 0.6, he/she gets access to the network and successfully bypasses the \mbox{ELECTRON}.  We also notice great variations into {\it office} communities, but they are more abrupt in {\it residence} community, where the base rate $a$ is higher. Thus, that behavior points out that higher base rates do not ensure higher detection, and that such communities are more sensible to attacks due to their fast trust convergence. Despite  these  worst results than to the other communities, {\it residence} and {\it office} communities had DR around 79.6\% and 90.6\%, respectively. Although changes in the attacker behavior diminish the \mbox{ELECTRON} performance, mainly due to the detection engine is based on trust and sociability, even so we verify an increase of DR.

The comparative analysis with \mbox{SA$^{2}$CI} reveals that the \mbox{ELECTRON} performance against Sybil attackers is better in scenarios with 100 and 150 nodes. In scenarios with 200 nodes, the cryptography-based mechanism obtains better DR than those provided by the community-based mechanism, and  that behavior is related to the number of nodes collaborating with opinions in the trust building process in \mbox{ELECTRON}. The greater the number of nodes, the more difficult is the consensus convergence into the established communities. On the other hand, environments with fewer devices enable more consistent opinions, and thus the trust management based on social relations is more effective in  supporting Sybil attack detection. These environments have also exhibited a more stable behavior for all communities.

\begin{figure}[!htbp]
    \centering
    \includegraphics[width=60mm]{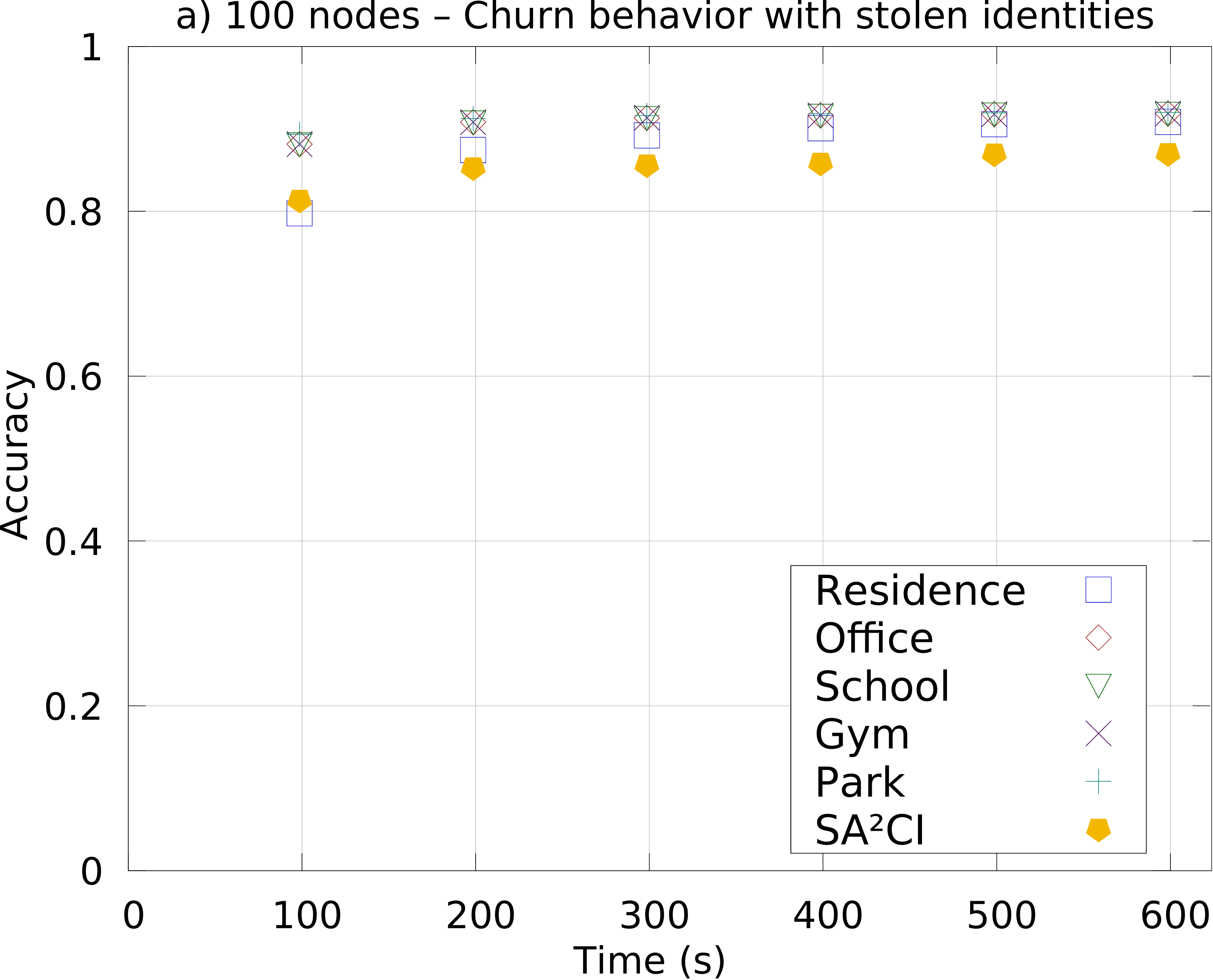}
    ~
    ~
    \includegraphics[width=60mm]{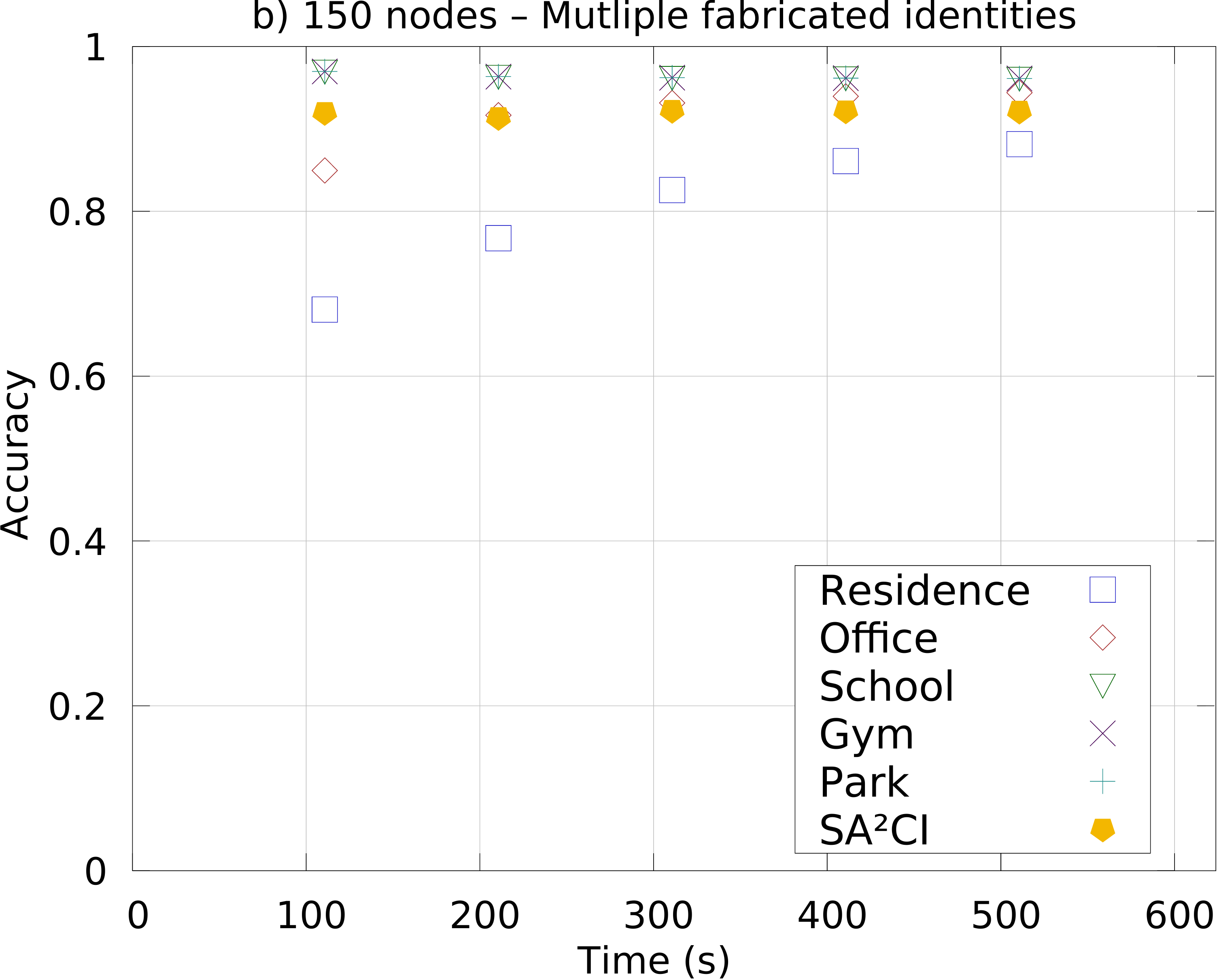}
    \caption{Accuracy in attacks detection}
    \label{fig:accuracyChurnStolen}
\end{figure} 

Figure~\ref{fig:accuracyChurnStolen}a presents the \mbox{ELECTRON} \textit{Accuracy} (ACC) in detecting Sybil attacks with {\it churn} behavior and stolen identities in the scenario with 100 nodes. We verify that the values of ACC (Figure~\ref{fig:accuracyChurnStolen}a) and DR (Figure~\ref{fig:detectedRate}b) had a similar trend with lower values in the {\it residence} community, while the other environments show almost identical values. This behavior also shows up in the scenario with 150 nodes and Sybil attacks employing multiple fabricated identities (Figure~\ref{fig:accuracyChurnStolen}b), in which {\it residence} and {\it office} communities show lowers results than others, while {\it school}, {\it academy}, and {\it park} communities present best values (i.e., values closer to 1). This means that greater amount of nodes disseminating recommendation collaborates to a better  accuracy detection, as the social trust becomes stable. Despite \mbox{SA$^{2}$CI} has achieved higher accuracy in {\it residence} and {\it office} communities, \mbox{ELECTRON} got better values in the other communities that are more exposed to attackers, as well as small and medium sized networks.

\begin{figure}[!htbp]
    \centering
    \includegraphics[width=60mm]{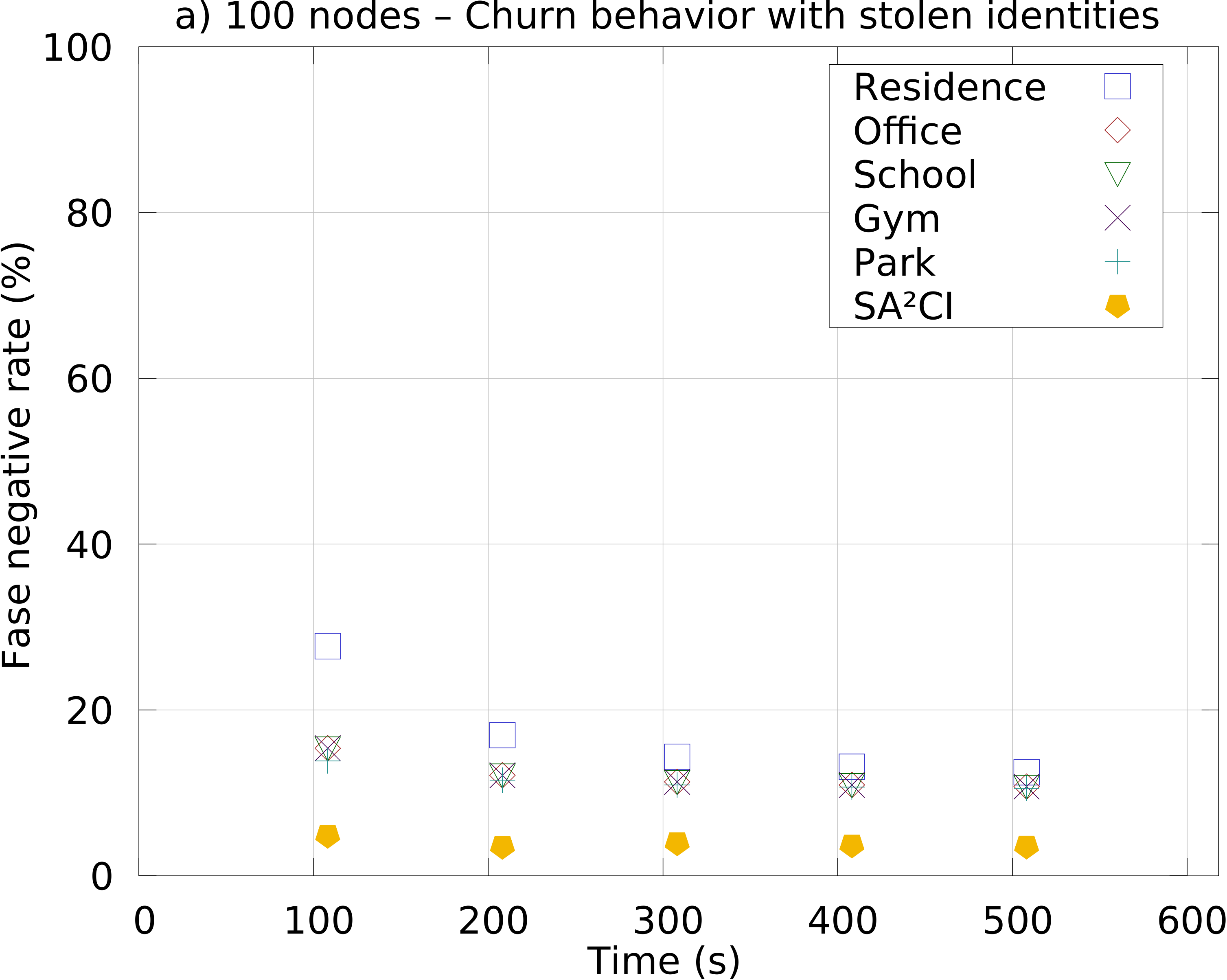}
    ~
    ~
    \includegraphics[width=60mm]{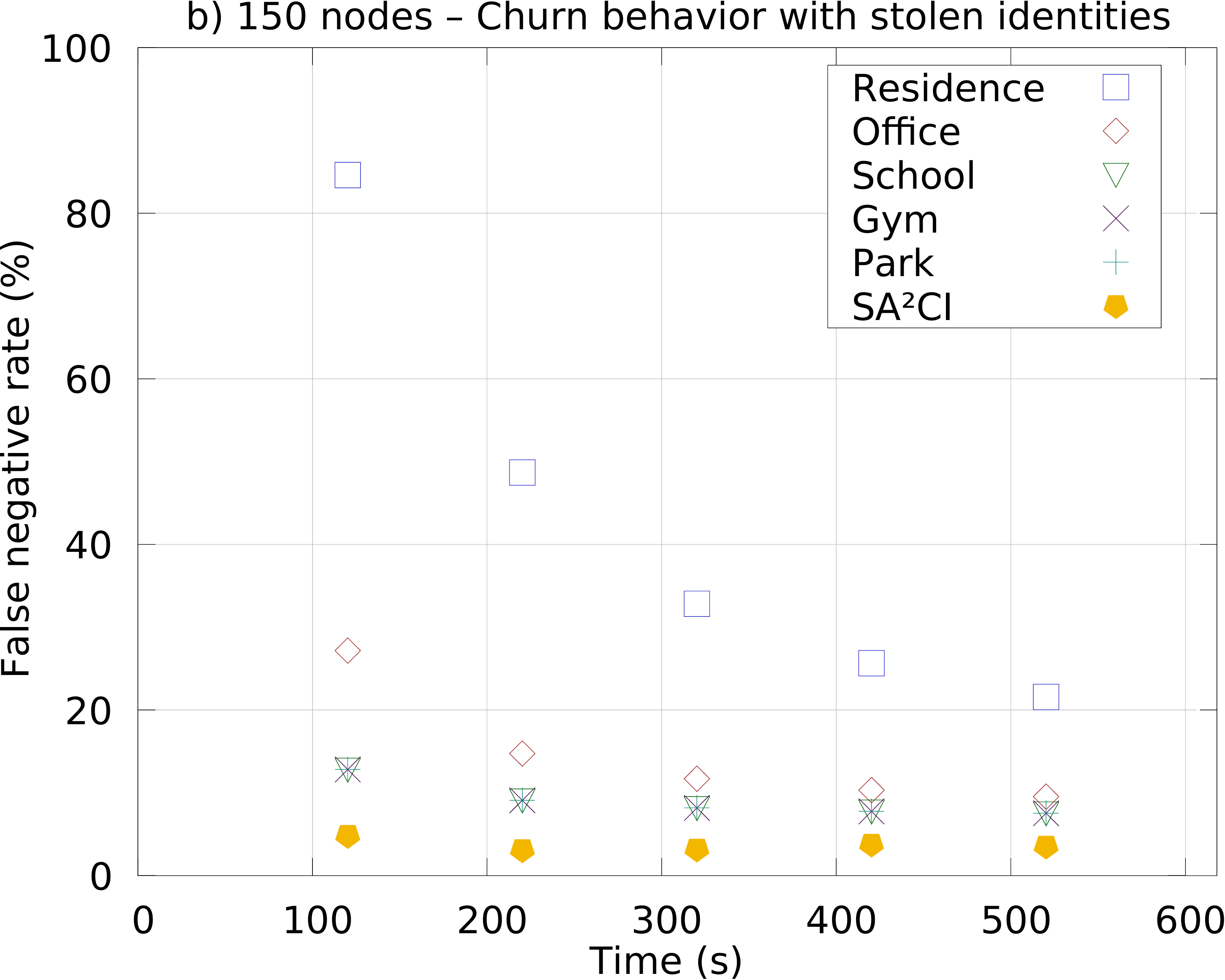}
    \caption{\textit{False Negative Rates in Sybil} attacks detection}
    \label{fig:falseNegChurnStolen}
\end{figure}  

Figure~\ref{fig:falseNegChurnStolen} exhibits the values of \textit{False Negative Rates} (FN) earned for both systems in  scenarios with 100 and 150 nodes. As FNs represent attackers classified as legitimate nodes, they are hence treated as a threat. In contrast,  FPs are legitimate nodes classified as attackers and, thus they are prevented from authenticating in the network. We can verify the efficiency of both systems in detecting Sybil attacks by comparing FN and DR (Figure~\ref{fig:detectedRate}) in all environments and in all established social relations, in order to demonstrate their operation diversity. Certain attackers were set up to simulate a legitimate node behavior, in order to circumvent the mechanism and achieve a greater trust. As shows Figure~\ref{fig:falseNegChurnStolen}(b), \mbox{ELECTRON} has achieved up to 90\% of FN in {\it residence} community in a scenario with 150 nodes against almost 30\% in a scenario with 100 nodes. We note that the identification of the attackers at the beginning of the simulation is a challenge, as the interactions are still insufficient to allow a robust trust construction. \mbox{ELECTRON} presents better results as devices trust evolve and it has obtained 22\% of FN for {\it residence} community and around 10\% for others close to the end of simulation in scenario with 150 nodes (Figure~\ref{fig:falseNegChurnStolen}(b)). \mbox{SA$^{2}$CI} has obtained lower values to FN, about 5\% in both scenarios, because it is less susceptible to changes in the attackers behavior when compared to \mbox{ELECTRON}.

\begin{figure}
    \centering
    \includegraphics[width=60mm]{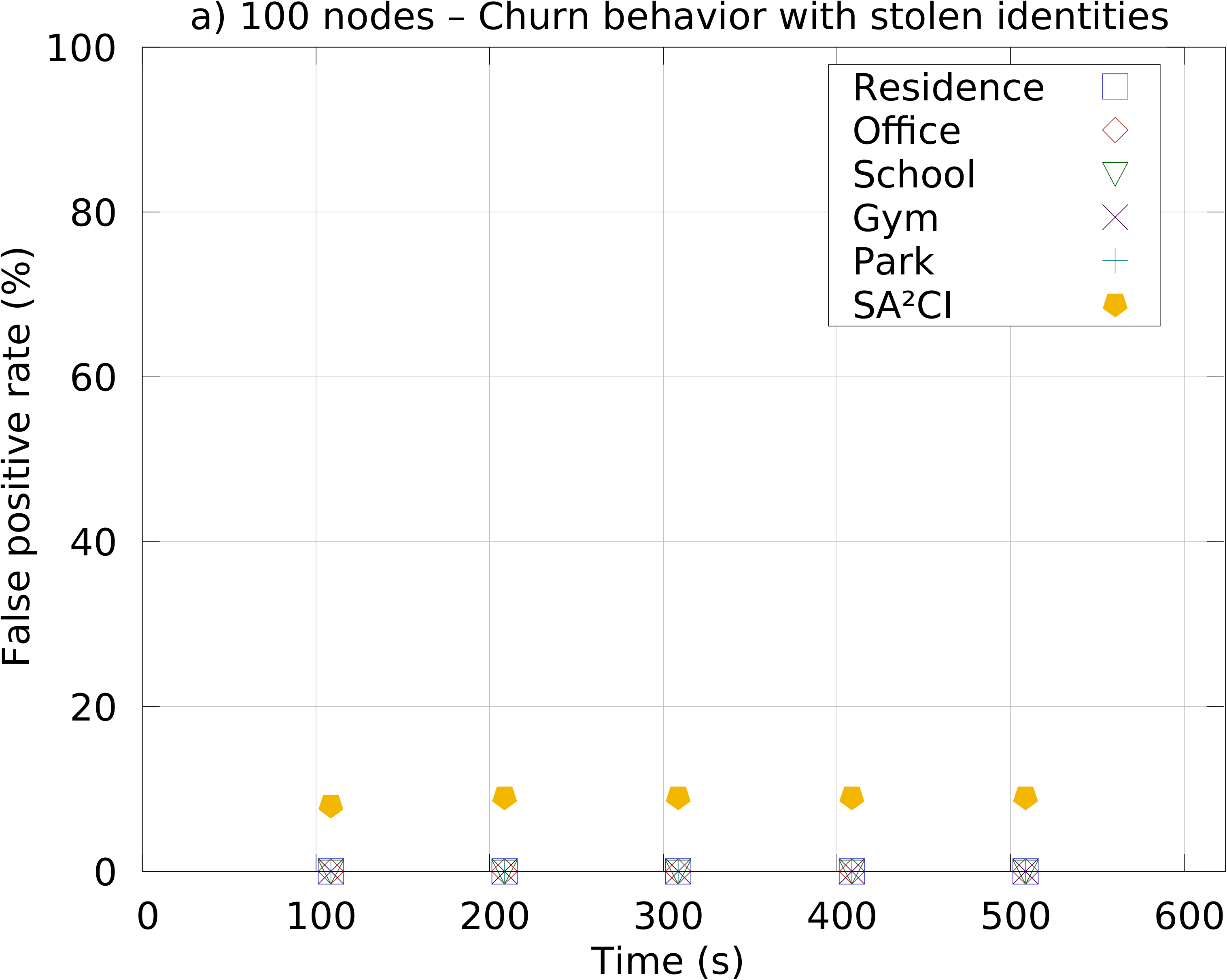}
    ~
    ~
    \includegraphics[width=60mm]{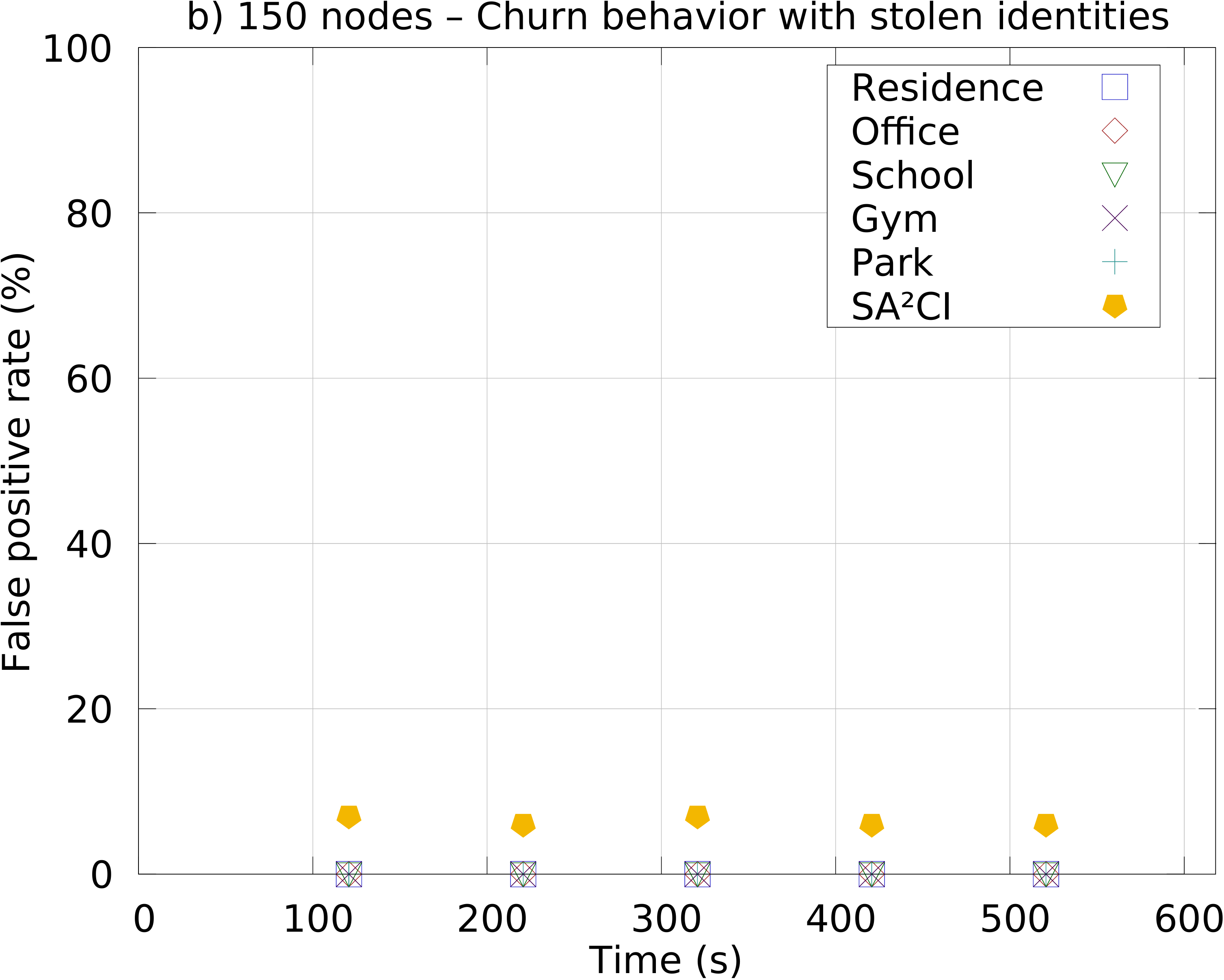}
    \caption{False positive rates in attacks detection}
    \label{fig:falsePosChurnStolen}
\end{figure}

Figure~\ref{fig:falsePosChurnStolen} exhibits the values of \textit{False Positive Rates} (FP) earned for \mbox{ELECTRON} and \mbox{SA$^{2}$CI} for scenarios with 100 and 150 nodes. There were no false positives in the \mbox{ELECTRON} classification, demonstrating that it  keeps stable levels of identification of legitimate nodes. Case such legitimate nodes came to act maliciously, \mbox{ELECTRON} would classify them as attackers. On other hand, \mbox{SA$^{2}$CI} had about 11.4\% of false positives, since attackers with a {\it churn} behavior imposes greater difficulty for detection in reason of the manner as \mbox{SA$^{2}$CI} forms nodes association and disassociation, which impairs the distinction between attacking nodes and legitimate ones. Hence, when analyzing FN and FP together, we note that they are similar for most communities of \mbox{ELECTRON} and \mbox{SA$^{2}$CI}.

\section{Conclusion and future works}\label{sec:concl}

This work presented \mbox{ELECTRON}, an access control mechanism for IoT networks based on devices social trust to protect against Sybil attackers. Through the social similarity between devices, boosted by context perception, \mbox{ELECTRON} clusters devices into communities and evaluates their social trust, strengthening the reliability between legitimate devices and their resilience against access attempts by Sybil attackers. Simulations evaluated the effectiveness of \mbox{ELECTRON}, and the results has demonstrated its ability in controlling devices access to the established network under Sybil attacks. \mbox{ELECTRON} prevented Sybil attackers to access the network by devices sociability and their similarity into communities, avoiding these attackers from launching other kinds of attacks and compromise users' data privacy. \mbox{ELECTRON} also achieved high accuracy and low false positive rates by social relations provided by device interactions and their similarity into available communities. The \mbox{ELECTRON} adaptability and flexibility enable its deployment in distinct environments and communities. Future works encompass investigating issues associated to the \mbox{ELECTRON} performance under packet losses and data collisions, besides applying fuzzy logic to improve devices authentication.

\section*{Acknowledgments}

We would like to acknowledge the support of the Brazilian Agency CNPq - grants \#309238/2017-0, \#436649/2018-7
and \#313641/2020-0.

\subsection*{Conflict of interest}

The authors declare no potential conflict of interests.

\bibliography{electron}

\clearpage

\section*{Author Biography}

\noindent
{\textbf{Gustavo Henrique Carvalho de Oliveira} received the B.E. degree from Federal Technological University of Paraná, Brazil, and is Master M.S. in Informatics from Federal University of Parana (UFPR). He is now working as system analyst. His research interests are data security, Internet of Things (IoT), Social Internet of Things (SIoT), access control and communication machine to machine (M2M).}

\vspace{8.0pt}

\noindent
{\textbf{Agnaldo de Souza Batista} received the B.E. degree from Catholic University of Pelotas, Brazil, and is Master M.S. in Informatics from Federal University of Parana (UFPR). He is now pursuing his Ph.D. degree in computer science at UFPR. His research interests are robust systems, data security, wireless networking, Internet of Things (IoT), \mbox{e-health}, and management of critical events. Agnaldo is member of Brazilian Computer Society (SBC).}

\vspace{8.0pt}

\noindent
{\textbf{Michele Nogueira} is a professor of computer science at Federal University of Minas Gerais, where she has been since 2010. She received her doctorate in computer science from the University Pierre et Marie Curie – Sorbonne Université, Laboratoire d’Informatique de Paris VI (LIP6) in 2009. She was a Visiting Researcher at Georgia Institute Technology (GeorgiaTech) and a Visiting Professor at University Paul Sabatier in 2009 and 2013, respectively. Her research interests include wireless networks, security and dependability. She has been a recipient of Academic Scholarships from Brazilian Government on her undergraduate and graduate years, and of international grants such as from the ACM SIGCOMM Geodiversity program. She is also Associate Technical Editor for the IEEE Communications Magazine and the Journal of Network and Systems Management.}

\vspace{8.0pt}

\noindent
{\textbf{Aldri L. dos Santos} is a professor of the Department of Computer Science  at Federal University of Minas Gerais (UFMG). Aldri is PhD in Computer Science from the Federal University of Minas Gerais, Master in Informatics and Bachelor of Computer Science at UFPR. Aldri is working in the following research areas: network management, fault tolerance, security, data dissemination, wireless ad hoc networks and sensor networks. He is leader of the research group (Wireless and Advanced Networks). Aldri has also acted as reviewer for publications as IEEE ComMag, IEEE ComNet, ComCom, IEEE Communications Surveys and Tutorials, IEEE eTNSM, JNSM, Ad hoc Networks. Aldri has served as member of the technical committee of security information and IEEE Communication Society Communication (ComSoc).}

\end{document}